\documentclass[pdflatex,sn-mathphys-num]{sn-jnl}%
\pdfoutput=1

\usepackage{graphicx}%
\usepackage{multirow}%
\usepackage{amsmath,amssymb,amsfonts}%
\usepackage{amsthm}%
\usepackage{mathrsfs}%
\usepackage[title]{appendix}%
\usepackage{xcolor}%
\usepackage{textcomp}%
\usepackage{manyfoot}%
\usepackage{booktabs}%

\let\orcidlogo\undefined %
\usepackage{orcidlink}
\usepackage{tikz}
\usetikzlibrary{quantikz}
\usepackage{physics}
\usepackage{float}
\usepackage{subcaption}
\usepackage{mathptmx}
\usepackage{microtype}

\newcommand{\prsuc}{\text{Pr}_\text{success}}

\theoremstyle{thmstyleone}%
\newtheorem{theorem}{Theorem}%
\newtheorem{proposition}[theorem]{Proposition}%

\begin{document}

\title[Noise Thresholds for Quantum Advantage in the Bernstein--Vazirani and Deutsch--Jozsa Algorithms]{Noise Thresholds for Quantum Advantage in the Bernstein--Vazirani and Deutsch--Jozsa Algorithms}

\author[1]{\fnm{Muhammad} \sur{Faizan}\,\orcidlink{0009-0000-1527-5275}}

\author*[2]{\fnm{Muhammad} \sur{Faryad}\,\orcidlink{0000-0003-1542-831X}}\email{muhammad.faryad@lums.edu.pk}

\affil[1]{\orgdiv{Department of Physics \& Astronomy}, \orgname{University of Nebraska--Lincoln}, \orgaddress{\city{Lincoln}, \postcode{68588}, \state{Nebraska}, \country{USA}}}

\affil*[2]{\orgdiv{Department of Physics}, \orgname{Lahore University of Management Sciences}, \orgaddress{\city{Lahore}, \postcode{54792}, \country{Pakistan}}}

\abstract{We analytically investigate the robustness of the Bernstein--Vazirani and Deutsch--Jozsa algorithms against single-qubit Pauli noise, of which bit-flip, phase-flip, and depolarizing noise are special cases, using the density-matrix formalism. For a general Pauli channel acting on each qubit after every gate, we derive exact closed-form expressions for the success probability of a single run as a function of the error probabilities and the number of qubits $n$, and validate them against Qiskit simulations. The noise enters only through the probability $\varepsilon$ with which each output bit is flipped. Under bit-flip noise with error probability $p>\tfrac{1}{2}$ the most probable outcome of the Bernstein--Vazirani algorithm is the complement of the correct one, which classical post-processing corrects, whereas depolarizing and phase-flip noise never bias the output toward a wrong outcome. For the Deutsch--Jozsa algorithm, the success probability is the average over constant and balanced oracles, and the results are given for the single-variable balanced oracle, which is shown to be the least robust of all balanced oracles for $\varepsilon<\tfrac{1}{2}$. For a single run, the tolerable error probability of both algorithms decreases as $1/n$. We then determine the minimum number of repetitions required for a given confidence and compare it with the number of queries of the classical algorithms. The threshold error probability below which the noisy Bernstein--Vazirani algorithm retains its query advantage increases with $n$, whereas that of the Deutsch--Jozsa algorithm with its standard readout decreases as $1/n$. This contrast indicates that algorithms whose answer is encoded in many independently decodable output bits are considerably more robust to noise than algorithms whose verdict depends on a single global outcome.}

\keywords{Bernstein--Vazirani Algorithm, Deutsch--Jozsa Algorithm, Quantum Algorithms, Quantum Noise, Quantum Computing, Quantum Information, Quantum Channels}

\maketitle

\tableofcontents

\section{Introduction}\label{sec:intro}

Quantum computation can solve certain classes of problems far more efficiently than classical computation. The Bernstein--Vazirani algorithm, for instance, exploits quantum parallelism to determine a hidden bit string with a single oracle query~\cite{bvalgo}, while Deutsch's algorithm provides one of the earliest demonstrations of quantum speedup~\cite{deutsch1992rapid}. Grover's algorithm achieves a quadratic speedup for unstructured search~\cite{grover}, and Shor's algorithm factors large integers in polynomial time~\cite{shor}. Together, these results show that quantum algorithms can outperform their classical counterparts on important computational tasks. For instance, a quantum advantage has been demonstrated experimentally on superconducting and photonic processors~\cite{arute,zhong,outperform}, with evidence of utility before fault tolerance~\cite{kim_utility}. Quantum advantage has also been proved theoretically for constant-depth circuits \cite{bravyi2018,bravyi2020NoisyShallowCircuits}. Near-term devices are also being explored for hybrid quantum--classical applications, such as quantum generative models and classifiers based on parameterized quantum circuits~\cite{lou_qucongan,liu_hybrid,lou_quanflex}.  However, noise has been the central obstacle to the practical implementation of quantum algorithms \cite{endo2021Hybrid}, which arises from imperfect gates and measurement devices and from the unwanted coupling of qubits to their environment~\cite{preskill,krantz,bruzewicz}. Characterizing how such noise affects algorithmic performance is therefore essential.

A number of studies have examined how noise affects specific quantum algorithms. Grover's search, for instance, remains more efficient than any classical counterpart in the presence of small amounts of noise~\cite{groverinnoise}. Simulations of quantum phase estimation show that the error in the estimated eigenvalue grows exponentially with the single-qubit error probability and, in the low-noise regime, increases linearly with the number of qubits~\cite{qpenoisefaizanfaryad}. For the Bernstein--Vazirani algorithm specifically, numerical work indicates that it retains some robustness against glassy disorder in the Hadamard gates, particularly for shorter secret strings~\cite{effectofnoise}. For the closely related problem of learning a parity function from noisy examples, the number of quantum queries grows only logarithmically with the number of bits under depolarizing noise, whereas the classical problem is believed to be intractable~\cite{cross2015}; this advantage has been demonstrated on a superconducting processor~\cite{riste2017}.

In contrast to these largely numerical investigations, we develop an analytical treatment of the effect of noise on the Bernstein--Vazirani and Deutsch--Jozsa algorithms. We focus on these algorithms because their shallow, low-complexity circuits make them analytically tractable while remaining foundational examples of quantum advantage~\cite{BVFernandez}. Additionally, these algorithms have been used as algorithmic benchmarks to assess the performance of trapped-ion and superconducting quantum processors \cite{lubinski2023Benchmarks,debnath2016Programmable,linke2017Comparison,blinov2021CloudComparison,wright}. The Bernstein--Vazirani algorithm exhibits a clear oracle separation between the quantum and bounded-error classical models: a quantum processor recovers the secret string in a single oracle query, whereas any bounded-error classical algorithm requires $\Omega(n)$ queries~\cite{qsimmdpi}. The closely related Deutsch--Jozsa algorithm, by comparison, attains an exponential advantage only over deterministic classical methods, since probabilistic classical algorithms solve it in constant time~\cite{djcontrast}.

As simple and representative models of hardware noise, we consider single-qubit Pauli channels, in which a qubit undergoes an $X$, $Y$, or $Z$ error with given probabilities. The \textit{bit-flip}, \textit{phase-flip}, and \textit{depolarizing} channels are special cases. These are among the most widely used error models and provide a standard, analytically tractable framework for analyzing how distinct error mechanisms act at the circuit level. The bit-flip model is useful for assessing algorithmic reliability because even a single bit-flip error can lead to an incorrect verdict about whether the function is constant or balanced in the Deutsch–Jozsa algorithm or corrupt the hidden string recovered by the Bernstein–Vazirani algorithm. A simplified bit-flip model approximately reproduced Deutsch--Jozsa output statistics on IBM Yorktown, supporting its use in error-mitigation analysis \cite{orrell2021}. The phase-flip channel describes pure dephasing, and the depolarizing channel, in particular, captures the average behavior of noise in large circuits containing many qubits and gates~\cite{mitigdep}. For gate- and time-independent errors, averaging over random Clifford gates reduces the noise to a depolarizing channel, which underlies randomized benchmarking~\cite{magesan}. Global depolarizing channels have also been used to describe and mitigate noise in deep circuits on superconducting processors~\cite{vovrosh}, and simulations based on a simple depolarizing error model agree with error-correction experiments on such processors~\cite{google2021}. Such Pauli noise is a central concern in quantum information because it degrades quantum coherence and, with it, the reliability of quantum computation~\cite{neilson}; understanding and mitigating it is therefore essential for dependable quantum information processing~\cite{Basit_2017,mitigdep}.

Real quantum hardware exhibits noise that is richer than this idealization: errors can be spatially and temporally correlated, as observed in cosmic-ray-induced bursts~\cite{mcewen} and long-range charge fluctuations~\cite{wilen} in superconducting qubits, and in the noise learned on multi-qubit devices~\cite{harper}, and the coupling of qubits to their environment is in general non-Markovian~\cite{devega}. Energy relaxation, moreover, is described by the amplitude-damping channel, which is not a Pauli channel. We nonetheless adopt independent Pauli channels as a first-order description, since randomized compiling converts general gate errors into Pauli noise~\cite{wallman}.

The structure of the paper is as follows. In Sections~\ref{sec:bv} and \ref{sec:dj} we present the Bernstein--Vazirani and Deutsch--Jozsa algorithms, respectively, in the density-matrix formalism and derive their success probabilities under a general Pauli channel, with bit-flip, phase-flip, and depolarizing noise as special cases, together with the number of repetitions required for a given confidence. In Section~\ref{sec:results} we compare the analytical results with Qiskit simulations and examine the dependence of the success probability on the error probability and the number of qubits, and the number of repetitions required for a given confidence, which we compare with the number of queries of the classical algorithms. A brief conclusion is presented in Section~\ref{sec:conclusions}. Appendix~\ref{app:deriv} contains the derivations of the success probabilities, and Appendix~\ref{app:nonuniform} treats qubit-dependent error probabilities.

\section{Bernstein--Vazirani Algorithm}\label{sec:bv}

The Bernstein--Vazirani algorithm determines a hidden binary string $\mathbf s \in \{0,1\}^n$ encoded in a Boolean function $f \colon \{0,1\}^n \to \{0,1\}$ of the form $f(\mathbf x) = \mathbf s \cdot \mathbf x \bmod 2$, where $\mathbf x \in \{0,1\}^n$ is the input and $\mathbf s \cdot \mathbf x = \bigoplus_{i=1}^n s_i x_i$ denotes the bitwise inner product taken modulo~2. 

Whereas a classical algorithm requires $n$ queries, one for each bit of $\mathbf s$, the Bernstein--Vazirani algorithm accomplishes this in a single query by exploiting quantum parallelism and superposition. The corresponding noiseless circuit is shown in Fig.~\ref{BV_circuit}.

\begin{figure}[H]
	\centering
	\begin{quantikz}
		\lstick{$\op{0}{0}$} & \gate{H}\slice{$\rho_1$} & \gate[4,nwires=3]{~~~~U_f~~~~}\slice{$\rho_2$} & \gate{H}\slice{$\rho_3$} & \meter{} \\
		\lstick{$\op{0}{0}$} & \gate{H} & & \gate{H} & \meter{} \\
		   & \vdots & \vdots & \vdots & \vdots \\
		\lstick{$\op{0}{0}$} & \gate{H} & & \gate{H} & \meter{}
	\end{quantikz}
	\caption{Circuit diagram of the noiseless Bernstein--Vazirani algorithm.}
	\label{BV_circuit}
\end{figure}

\subsection{Density-matrix formalism}\label{sec:bv_ideal}
The register is initialized in the state $\rho_0$; a layer of Hadamard gates $H^{\otimes n}$ then acts, followed by the oracle $U_f$ that implements $f(\mathbf x)$ and a second Hadamard layer, after which a measurement in the computational basis reveals the hidden string $\mathbf s$. Writing $N = 2^n$ and, for brevity, $H \equiv H^{\otimes n}$, the density matrix evolves through the circuit as
\begin{align}
	\rho_0  &= \op{\mathbf{0}}{\mathbf{0}}, \qquad \ket{\mathbf 0} = \ket{0}^{\otimes n} \label{eq1} \\
	\rho_1 & = H\op{\mathbf{0}}{\mathbf{0}}H = \frac{1}{N} \sum_{x,y=0}^{N-1}\op{x}{y} \label{rho1} \\
	\rho_2 & = U_f \rho_1 U_f = \frac{1}{N} \sum_{x,y=0}^{N-1} (-1)^{f(x)+f(y)} \op{x}{y} \label{rho2} \\
	\rho_3 & = H \rho_2 H = \sum_{a,b=0}^{N-1} C_{ab} \op{a}{b} \label{rho3}
\end{align}
where the coefficients are
\begin{equation}
	C_{ab} = \frac{1}{N^2} \sum_{x,y=0}^{N-1}(-1)^{f(x)+\mathbf{a}\cdot \mathbf x}\,(-1)^{f(y) + \mathbf{b}\cdot \mathbf y}.
\end{equation}
The hidden string is identified from the most probable outcome of a measurement of $\rho_3$. The probability of obtaining the computational basis state $\ket{\mathbf x}$ is
\begin{align}
	\Pr(\ket{\mathbf x}) = \mel{\mathbf x}{\rho_3}{\mathbf x}.
\end{align}
The algorithm succeeds when this probability is maximal at $\ket{\mathbf x} = \ket{\mathbf s}$. In the noiseless case the maximum equals unity,
\begin{align}
	\max_{\ket{\mathbf x}} \mel{\mathbf x}{\rho_3}{\mathbf x} = 1, \qquad \text{attained at } \ket{\mathbf x}=\ket{\mathbf s}.
\end{align}

\begin{proposition}
	Let $f(\mathbf x) = \mathbf s \cdot \mathbf x \bmod 2$ be a Boolean function for $\mathbf s \in \{0,1\}^n$. Then the oracle $U_f$ defined by
	\begin{equation}
		U_f \ket{\mathbf x} = (-1)^{f(\mathbf x)} \ket{\mathbf x}
	\end{equation}
	can be written as a tensor product of single-qubit unitaries:
	\begin{equation}
		U_f = \bigotimes_{i=1}^n Z^{s_i},
	\end{equation}
	where $Z$ is the Pauli-$Z$ gate and $Z^{s_i}$ denotes $Z$ if $s_i = 1$ and the identity $\mathbb{I}$ otherwise.
\end{proposition}

\begin{proof}
	Since $f(\mathbf x) = \mathbf s \cdot \mathbf x = \bigoplus_{i=1}^n s_i x_i$, we have
	\begin{equation}
		(-1)^{f(\mathbf x)} = \prod_{i=1}^n (-1)^{s_i x_i}.		
	\end{equation}
	Thus, the oracle acts as
	\begin{equation}
		U_f \ket{\mathbf x} = \left( \prod_{i=1}^n (-1)^{s_i x_i} \right) \ket{\mathbf x}.		
	\end{equation}
	Each factor $(-1)^{s_i x_i}$ corresponds to applying a Pauli-$Z$ gate to the $i$\textsuperscript{th} qubit if $s_i = 1$ and doing nothing otherwise. Hence, define
	\begin{equation}
		U_i = \begin{cases}
			Z & \text{if } s_i = 1, \\
			\mathbb{I} & \text{if } s_i = 0,
		\end{cases}
	\end{equation}
	and the overall oracle is the tensor product
	\begin{equation}
		U_f = \bigotimes_{i=1}^n U_i = \bigotimes_{i=1}^n Z^{s_i}. \qedhere
	\end{equation}
\end{proof}
The corresponding quantum circuit is shown in Fig.~\ref{oracle-decomposition}.
	
\begin{figure}[H]
	\centering
	\begin{quantikz}[row sep=0.5mm]
	& \gate[4,nwires=3]{~~~U_f~~~}	& \qw \\
	& & \qw \\
	\vdots & & \vdots  \\
	& & \qw
	\end{quantikz} \quad $\equiv$ \quad \begin{quantikz}[row sep=1mm]
	& \gate{Z^{s_1}} & \qw \\
	& \gate{Z^{s_2}} & \qw \\
	\vdots & \vdots & \vdots \\
	& \gate{Z^{s_n}} & \qw
	\end{quantikz}
	\caption{Circuit representation of the oracle $U_f = \bigotimes_{i=1}^n Z^{s_i}$: each qubit receives a $Z$ gate if the corresponding bit $s_i = 1$ and the identity otherwise. This implements the phase oracle for the Boolean function $f(\mathbf{x}) = \mathbf{s} \cdot \mathbf{x} \bmod 2$.}
	\label{oracle-decomposition}
\end{figure}

\subsection{Bernstein--Vazirani algorithm in the presence of noise}\label{sec:bv_noise}

We now assume that noise acts independently on each qubit after every gate. For clarity, the same error probabilities are taken for every qubit. Since the error rates in current devices are small, of the order of $10^{-3}$~\cite{ballance2016,foxen2020}, qubit-dependent error probabilities change the results by a negligible amount, so that the error probabilities may be read as their mean values (see Appendix~\ref{app:nonuniform}). The circuit for the $i$\textsuperscript{th} qubit is shown in Fig.~\ref{BV_noisy}.

\begin{figure}[H]
	\centering
	\begin{tikzpicture}
		\node[scale=1]{
			\begin{quantikz}
				\lstick{$\op{0}{0}$}  & \gate{H}\slice{$\phi^i_1$} & \gate[style={fill=red!20}]{\mathcal{E}}\slice{$\phi^i_2$} & \gate{Z^{s_i}}\slice{$\phi^i_3$} & \gate[style={fill=red!20}]{\mathcal{E}}\slice{$\phi^i_4$} & \gate{H}\slice{$\phi^i_5$} & \gate[style={fill=red!20}]{\mathcal{E}}\slice{$\phi^i_6$} & \meter{}
			\end{quantikz}
		};
	\end{tikzpicture}
	\caption{Circuit diagram of the Bernstein--Vazirani algorithm with noise acting independently on each qubit after every gate.}
	\label{BV_noisy}
\end{figure}
We derive a closed-form expression for the success probability, i.e., the probability of correctly identifying the secret string $\ket{\mathbf s}$, for a general single-qubit Pauli channel. Bit-flip, phase-flip, and depolarizing noise then follow as special cases. The derivations are presented in Appendix~\ref{app:deriv}.

\subsubsection{Pauli noise}\label{sec:bv_pauli}
A single-qubit Pauli channel applies the Pauli operators $X$, $Y$, and $Z$ with probabilities $p_x$, $p_y$, and $p_z$, respectively, and leaves the qubit unchanged otherwise:
\begin{equation}\label{paulidef}
	\mathcal{E}(\rho) = (1-p_x-p_y-p_z)\,\rho + p_x\,X\rho X + p_y\,Y\rho Y + p_z\,Z\rho Z.
\end{equation}
When this channel acts on each qubit after every gate (Fig.~\ref{BV_noisy}), each measured bit is flipped with probability
\begin{equation}\label{eps_def}
	\varepsilon = \frac{1}{2}\left(1-r_x^{2}\,r_z\right), \qquad r_x = 1-2(p_y+p_z), \qquad r_z = 1-2(p_x+p_y),
\end{equation}
independently of the other bits, and the secret string $\mathbf s$ is obtained with probability
\begin{equation}\label{prsuc_pauli}
	\prsuc = (1-\varepsilon)^n.
\end{equation}
Here $r_x$ and $r_z$ are the factors by which the channel scales the $X$ and $Z$ components of the Bloch vector of a qubit. The factor $r_x$ enters twice because the qubit lies along the $x$ axis of the Bloch sphere during the first two noise layers, and $r_z$ enters once because the qubit lies along the $z$ axis during the third. The noise thus enters only through the bit-error probability $\varepsilon$. For $\varepsilon < \tfrac{1}{2}$, $\ket{\mathbf s}$ is the most probable outcome, whereas for $\varepsilon > \tfrac{1}{2}$ the most probable outcome is the complementary string $\ket{\bar{\mathbf s}}$, from which $\mathbf s$ can be retrieved by classical post-processing.

Since the noise acts independently on each qubit, each output bit is correct with probability $1-\varepsilon$, independently of the other bits, so that the reliability can be increased by repeating the algorithm. Suppose that the circuit is run an odd number $m$ of times and that each bit is decided by majority vote, followed by complementation if $\varepsilon>\tfrac{1}{2}$. In each run a given bit is then correct with probability
\begin{equation}\label{q_bit}
	q = \begin{cases}
		1-\varepsilon & \text{for } \varepsilon<\tfrac{1}{2},\\
		\varepsilon & \text{for } \varepsilon>\tfrac{1}{2} \text{ (after complementation).}
	\end{cases}
\end{equation}
Let $S_i$ denote the number of runs in which the $i$\textsuperscript{th} bit is correct. The majority vote on this bit is correct if
\begin{equation}
	S_i \geq \frac{m+1}{2},
\end{equation}
i.e., if the bit is correct in more than half of the runs. Since the runs are independent, $S_i$ follows a binomial distribution, and the vote is correct with probability
\begin{equation}\label{bit_majority}
	P_{\rm bit}(m) = \sum_{k=(m+1)/2}^{m}\binom{m}{k}\,q^{k}(1-q)^{m-k}.
\end{equation}
The $n$ bits are independent, so that the secret string $\mathbf s$ is identified with probability
\begin{equation}\label{exact_bv}
	\prsuc(m) = \left[P_{\rm bit}(m)\right]^{n}.
\end{equation}
Let $\delta$ denote the largest failure probability that is acceptable in a given application, so that $1-\delta$ is the required confidence (e.g., $\delta=0.01$ for $99\%$ confidence). The required number of runs is the smallest $m$ for which the success probability reaches this confidence,
\begin{equation}\label{m_min_bv}
	m_{\min}(n,\varepsilon) = \min\left\{ m \text{ odd} : \prsuc(m) \geq 1-\delta \right\}.
\end{equation}

\subsubsection{Bit-flip, phase-flip, and depolarizing noise}\label{BVbitflip}
The bit-flip channel flips a qubit between $\ket{0}$ and $\ket{1}$ with probability $p$ and leaves it unchanged otherwise:
\begin{equation}\label{bitflipdef}
	\mathcal{E}(\rho) = (1-p)\,\rho + p\,X\rho X.
\end{equation}
It is the Pauli channel of Eq.~\eqref{paulidef} with $p_x=p$ and $p_y=p_z=0$, for which $r_x=1$, $r_z=1-2p$, and $\varepsilon=p$. Equation~\eqref{prsuc_pauli} then gives
\begin{equation}\label{prsuc_bf}
	\prsuc^{(X)} = (1-p)^n.
\end{equation}
Table~\ref{tab:repetitions} lists $m_{\min}$ for $\delta=0.01$, together with Qiskit simulations of the majority vote.

The simulated success probabilities in this and the following tables are obtained as follows. For a given $m$, one trial consists of $m$ runs of the noisy circuit followed by the decision rule, which is the bitwise majority vote here and the threshold rule of Sec.~\ref{sec:dj_pauli} for the Deutsch--Jozsa algorithm, and yields a single outcome. The trial is repeated $T=2000$ times, so that the circuit is executed $mT$ times in total, and the simulated success probability is the fraction of trials with a correct verdict. For success probabilities close to $0.99$, the standard error of this fraction is about $0.002$.

\begin{table}[!htbp]
	\centering
	\caption{Number of repetitions $m$ required to identify the secret string $\mathbf s$ with confidence $1-\delta=0.99$ under bit-flip noise, using a bitwise majority vote. $m_{\min}$ is the smallest odd $m$ for which the success probability of Eq.~\eqref{exact_bv} reaches $1-\delta$ [Eq.~\eqref{m_min_bv}]; the analytical and simulated success probabilities at $m_{\min}$ are also listed. Simulated values are obtained from $2000$ trials of $m$ runs each, as described in Sec.~\ref{BVbitflip}.}
	\label{tab:repetitions}
	\begin{tabular*}{\linewidth}{c@{\extracolsep{\fill}}c ccc}
		\toprule
		$n$ & $p$ & $m_{\min}$ & $\prsuc$ (analytical) & $\prsuc$ (Qiskit) \\
		\midrule
		10 & 0.1 & 9  & 0.9911 & 0.9885 \\
		10 & 0.3 & 55 & 0.9907 & 0.9915 \\
		20 & 0.1 & 11 & 0.9941 & 0.9955 \\
		20 & 0.3 & 63 & 0.9913 & 0.9935 \\
		30 & 0.1 & 11 & 0.9912 & 0.9925 \\
		30 & 0.3 & 67 & 0.9910 & 0.9895 \\
		
		\bottomrule
	\end{tabular*}
\end{table}

The phase-flip channel applies a $Z$ error with probability $p$,
\begin{equation}\label{phaseflipdef}
	\mathcal{E}(\rho) = (1-p)\,\rho + p\,Z\rho Z,
\end{equation}
and describes pure dephasing. It is the Pauli channel with $p_z=p$ and $p_x=p_y=0$, for which $r_x=1-2p$ and $r_z=1$, so that
\begin{equation}\label{prsuc_pf}
	\varepsilon = 2p(1-p), \qquad \prsuc^{(Z)} = \left[1-2p(1-p)\right]^n.
\end{equation}
Here $\varepsilon\leq\tfrac{1}{2}$, so that no complementation is required. Moreover, $\varepsilon$ is symmetric about $p=\tfrac{1}{2}$ and vanishes at $p=1$, where the phase flips of the first two noise layers cancel and the phase flip before the measurement does not change its outcome.

The depolarizing channel replaces an arbitrary $d$-dimensional state $\rho$ by the maximally mixed state $\mathbb I/d$ with probability $p$:
\begin{equation}\label{dep}
	\mathcal{E}(\rho) = (1-p)\,\rho + p\,\frac{\mathbb I}{d}.
\end{equation}
For a single qubit ($d = 2$) the channel reads $\mathcal{E}(\rho) = (1-p)\rho + \tfrac{p}{2}\mathbb{I}$. It is the Pauli channel with $p_x=p_y=p_z=p/4$, for which $r_x=r_z=1-p$ and $\varepsilon=\beta$, where
\begin{equation}\label{beta_def}
	\beta = \tfrac{1}{2}\left[1-(1-p)^3\right] = \tfrac{1}{2}\,p\,(p^2 - 3p + 3).
\end{equation}
Equation~\eqref{prsuc_pauli} then gives
\begin{equation}\label{prsuc_dep}
	\prsuc^{\text{dep}} = (1-\beta)^n = \left(\alpha + \beta \right)^n,
\end{equation}
with $\alpha = (1-p)^3$. Each output bit is thus correct with probability $\alpha+\beta=\tfrac{1}{2}\left[1+(1-p)^3\right]$ and wrong with probability $\beta$. Since $\beta<\tfrac{1}{2}$ for all $p<1$, no complementation is required, and the minimum number of runs follows from Eqs.~\eqref{bit_majority}--\eqref{m_min_bv} with
\begin{equation}
	q = \alpha+\beta = \frac{1+(1-p)^3}{2}.
\end{equation}
The corresponding values are compared with Qiskit simulations in Table~\ref{tab:rep_bv_dep}. Since $m_{\min}$ depends on the channel only through $\varepsilon$, Table~\ref{tab:repetitions} applies to any Pauli channel whose bit-error probability $\varepsilon$ equals the listed value of $p$.

\begin{table}[!htbp]
	\centering
	\caption{As Table~\ref{tab:repetitions}, but for depolarizing noise, for which $q=\alpha+\beta$ in Eq.~\eqref{bit_majority}. Simulated values are obtained from $2000$ trials of $m$ runs each, as described in Sec.~\ref{BVbitflip}.}
	\label{tab:rep_bv_dep}
	\begin{tabular*}{\linewidth}{c@{\extracolsep{\fill}}c ccc}
		\toprule
		$n$ & $p$ & $m_{\min}$ & $\prsuc$ (analytical) & $\prsuc$ (Qiskit) \\
		\midrule
		10 & 0.1 & 13 & 0.9932 & 0.9925 \\
		10 & 0.3 & 77 & 0.9908 & 0.9925 \\
		20 & 0.1 & 15 & 0.9940 & 0.9935 \\
		20 & 0.3 & 87 & 0.9907 & 0.9885 \\
		30 & 0.1 & 15 & 0.9911 & 0.9920 \\
		30 & 0.3 & 93 & 0.9907 & 0.9865 \\
		
		\bottomrule
	\end{tabular*}
\end{table}

The three channels differ qualitatively in how they limit performance. Under bit-flip noise, the most probable string is $\ket{\mathbf s}$ for $p<\tfrac{1}{2}$ and the complementary string $\ket{\bar{\mathbf s}}$ for $p>\tfrac{1}{2}$, from which $\mathbf s$ is recovered by complementing every measured bit. The algorithm therefore fails only at $p=\tfrac{1}{2}$, where each output bit is uniformly random and no information about $\mathbf s$ remains. Accordingly, the required number of repetitions $m_{\min}$ of Eq.~\eqref{m_min_bv} diverges as $p\to\tfrac{1}{2}$. The depolarizing channel, by contrast, drives the state toward the maximally mixed state without favoring any particular outcome; it therefore never biases the measurement away from $\ket{\mathbf s}$, although the success probability still decays with both $p$ and $n$. The phase-flip channel also never biases the measurement away from $\ket{\mathbf s}$, and its effect disappears again as $p\to1$.

\section{Deutsch--Jozsa Algorithm}\label{sec:dj}

The Deutsch--Jozsa algorithm determines whether the function $f(\mathbf x)$ implemented by the oracle $U_f$ is \textit{constant} or \textit{balanced}. Its structure is identical to that of the Bernstein--Vazirani algorithm, except that the oracle for a balanced function may be entangling. The final measurement reveals the nature of $f$: the function is constant if the output is $\ket{0}^{\otimes n}$ and balanced otherwise. We denote by $P_{\rm c}$ and $P_{\rm b}$ the probabilities of obtaining the outcome $\ket{\mathbf 0}$ when the function is constant and balanced, respectively. A single run then gives the correct verdict with probability $P_{\rm c}$ for a constant oracle and $1-P_{\rm b}$ for a balanced one. If the two cases are equally likely, then the success probability is the average of the probabilities of these two events,
\begin{equation}\label{dj_avg}
	\prsuc = \frac{1}{2}\left[P_{\rm c}+(1-P_{\rm b})\right] = \frac{1}{2}\left(1+\Delta\right), \qquad \Delta = P_{\rm c}-P_{\rm b}.
\end{equation} 
For a constant function the oracle reduces to a global phase, $U_f = \pm \mathbb{I}$, which is non-entangling, so the circuit and its success probability coincide with those of the Bernstein--Vazirani algorithm analyzed above. $P_{\rm c}$ is therefore given by Eq.~\eqref{prsuc_pauli}, and the probability $P_{\rm b}=\Pr(\ket{\mathbf 0})$ for the balanced case, for which the outcome $\ket{\mathbf 0}$ is the incorrect verdict, is derived in Appendix~\ref{app:deriv_dj}.

\subsection{Algorithm in the absence of noise}\label{sec:dj_ideal}
As in the Bernstein--Vazirani algorithm, the register is initialized in $\rho_0 = \op{\mathbf 0}{\mathbf 0}$; a Hadamard layer $H^{\otimes n}$, the oracle $U_f$, and a second Hadamard layer act in sequence before the final measurement. The density matrix evolves as
\begin{align}
	\rho_1 & = \frac{1}{N} \sum_{x,y=0}^{N-1} \op{x}{y}\\
	\rho_2 & = \frac{1}{N} \sum_{x,y=0}^{N-1} (-1)^{f(x) + f(y)} \op{x}{y}\\
	\rho_3 & = \sum_{a,b=0}^{N-1} C_{ab} \op{a}{b}\label{dj_rho3}
\end{align}
where $C_{ab}=\frac{1}{N^2}\sum_{x,y=0}^{N-1} (-1)^{f(x) + \mathbf a\cdot \mathbf x}\,(-1)^{f(y) + \mathbf b\cdot \mathbf y}$. Unlike in the Bernstein--Vazirani case, $\rho_2$ and $\rho_3$ need not be product states. The probability that $\rho_3$ lies in $\ket{0}^{\otimes n}$ is
\begin{align}
	\Pr(\ket{\mathbf 0}) & = \abs{\frac{1}{N} \sum_x (-1)^{f(x)}}^2 = \begin{cases}
		1 & f(x) \text{ is constant}, \\
		0 & f(x) \text{ is balanced}.
	\end{cases}
\end{align}

\begin{figure}[H]
	\centering
	\includegraphics[width=0.98\linewidth]{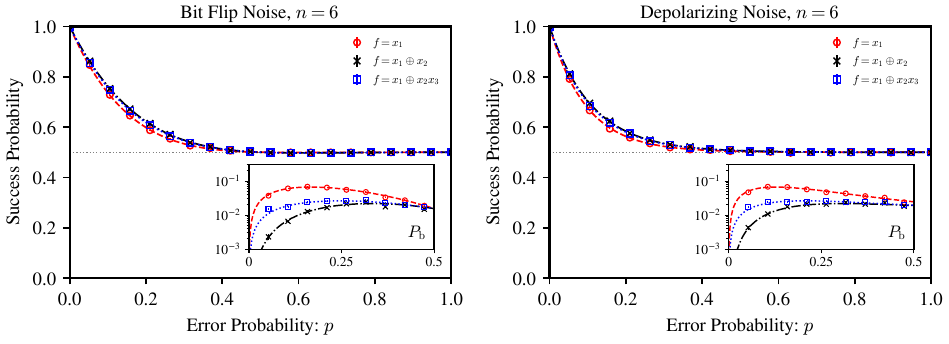}
	\caption{Success probability of the Deutsch--Jozsa algorithm for three balanced oracles at $n=6$ under bit-flip (left) and depolarizing (right) noise. \textbf{Symbols}: Qiskit simulations ($10^4$ shots per circuit; error bars are standard errors). \textbf{Lines}: analytical results [Eq.~\eqref{dj_avg} with the values of $P_{\rm b}$ given in the text]. The success probability is the average over a constant oracle and the balanced oracle indicated. The insets show the probability $P_{\rm b}$ of the incorrect outcome $\ket{\mathbf 0}$ for the balanced oracle on a logarithmic scale. For $\varepsilon<\tfrac{1}{2}$, i.e., for $p<\tfrac{1}{2}$ under bit-flip noise and for all $p<1$ under depolarizing noise, the single-variable oracle $f(\mathbf x)=x_1$ has the largest $P_{\rm b}$ and the lowest success probability. The dotted line marks the success probability $\tfrac{1}{2}$ of random guessing.}
	\label{fig:oracle_comparison}
\end{figure}

Throughout the noise analysis we take the representative single-variable balanced function $f(\mathbf x) = x_1$, whose phase oracle is a single $Z$ gate on the first qubit. We choose this oracle because, within the present model, it is the least robust of all balanced oracles. The verdict fails only if the noise maps the ideal output onto $\ket{\mathbf 0}$. For $f(\mathbf x)=x_1$ the ideal output is $\ket{10\cdots0}$, which a single bit flip converts into $\ket{\mathbf 0}$, whereas for other balanced functions more than one bit flip is required for part or all of the ideal output. Appendix~\ref{app:deriv_dj} makes this precise. For any balanced function, $P_{\rm b}$ is the average of $\varepsilon^{w}(1-\varepsilon)^{n-w}$ over the ideal output distribution, where $w\geq1$ is the Hamming weight of the output string and $\varepsilon$ is the probability that a measured bit is flipped [Eq.~\eqref{dj_spectrum}]. For $\varepsilon<\tfrac{1}{2}$ it is therefore largest when the ideal output has weight one, i.e., for a single-variable oracle. Figure~\ref{fig:oracle_comparison} illustrates this for $n=6$ by comparing $f(\mathbf x)=x_1$ with the linear oracle $f(\mathbf x)=x_1\oplus x_2$ and the entangling oracle $f(\mathbf x)=x_1\oplus x_2x_3$, for which $P_{\rm b}=\varepsilon^{2}(1-\varepsilon)^{n-2}$ and $P_{\rm b}=\tfrac{1}{4}\varepsilon(1-\varepsilon)^{n-3}$, respectively. For $\varepsilon<\tfrac{1}{2}$ the single-variable oracle has the largest $P_{\rm b}$ and hence the lowest success probability, defined in Eq.~\eqref{dj_avg}, and the three oracles become indistinguishable only where the success probability approaches $\tfrac{1}{2}$. The closed-form expressions obtained for $f(\mathbf x)=x_1$ therefore represent the least favorable case.

\subsection{Algorithm in the presence of noise}\label{sec:dj_noise}
\subsubsection{Pauli noise}\label{sec:dj_pauli}
The Pauli channel was defined in Eq.~\eqref{paulidef}. For a constant oracle the circuit coincides with the Bernstein--Vazirani circuit for $\mathbf s=\mathbf 0$, and for the balanced oracle $f(\mathbf x)=x_1$ it coincides with that for $\mathbf s=(1,0,\dots,0)$. Each measured bit is therefore flipped with the probability $\varepsilon$ of Eq.~\eqref{eps_def}, so that
\begin{equation}\label{dj_pc_pb}
	P_{\rm c} = (1-\varepsilon)^n, \qquad P_{\rm b} = \varepsilon(1-\varepsilon)^{n-1}.
\end{equation}
With these values, the success probability of Eq.~\eqref{dj_avg} is
\begin{equation}\label{dj_success_eps}
	\prsuc = \frac{1}{2}\left[1+\left(1-\varepsilon\right)^{n-1}\left(1-2\varepsilon\right)\right],
\end{equation}
which decreases from $1$ in the absence of noise to $\tfrac{1}{2}$, i.e., random guessing, at $\varepsilon=\tfrac{1}{2}$, where the noise makes the two cases indistinguishable. The corresponding expressions for an arbitrary balanced function are derived in Appendix~\ref{app:deriv_dj}.

In the presence of noise, a single run thus distinguishes a constant from a balanced oracle only probabilistically, so that the algorithm must be repeated. Suppose that the circuit is run $m$ times and let $M$ be the number of runs that yield $\ket{\mathbf 0}$. Since $P_{\rm c}>P_{\rm b}$ for $\varepsilon<\tfrac{1}{2}$, a constant oracle yields $\ket{\mathbf 0}$ more often than a balanced one. The oracle is therefore declared constant if $M>K$ and balanced otherwise, where $K$ is an integer threshold. Since $M$ follows a binomial distribution with parameter $P_{\rm c}$ or $P_{\rm b}$, the verdict is correct with probability
\begin{equation}\label{dj_const_bal}
	P_{\rm const}(m,K) = \sum_{k=K+1}^{m}\binom{m}{k}P_{\rm c}^{k}(1-P_{\rm c})^{m-k}, \qquad
	P_{\rm bal}(m,K) = \sum_{k=0}^{K}\binom{m}{k}P_{\rm b}^{k}(1-P_{\rm b})^{m-k}
\end{equation}
for a constant and a balanced oracle, respectively. We require the confidence $1-\delta$ for either type of oracle, in the same way as the error probability of the classical algorithm in Sec.~\ref{sec:repetitions} refers to its less favorable case. For known $\varepsilon$, the threshold $K$ is chosen to maximize the smaller of the two probabilities, so that
\begin{equation}\label{exact_dj}
	\prsuc(m) = \max_{0\leq K<m}\,\min\left\{P_{\rm const}(m,K),\; P_{\rm bal}(m,K)\right\}
\end{equation}
is the success probability of the repeated algorithm. Unlike Eq.~\eqref{dj_avg}, which is an average over the two types of oracle, Eq.~\eqref{exact_dj} refers to the less favorable one, so that the two do not coincide for $m=1$. The required number of runs $m_{\min}(n,\varepsilon)$ is the smallest $m$ for which Eq.~\eqref{exact_dj} is at least $1-\delta$. Because $K$ is an integer, Eq.~\eqref{exact_dj} is not monotonic in $m$; if more than $m_{\min}$ runs are available, the surplus can be discarded. Since $P_{\rm bal}$ decreases as $P_{\rm b}$ increases, and $f(\mathbf x)=x_1$ has the largest $P_{\rm b}$ of all balanced functions for $\varepsilon<\tfrac{1}{2}$, this number of runs, with the same threshold $K$, is sufficient for every balanced oracle.

\subsubsection{Bit-flip, phase-flip, and depolarizing noise}\label{sec:bf}
For the bit-flip channel of Eq.~\eqref{bitflipdef}, $\varepsilon=p$, so that
\begin{equation}\label{dj_pc_pb_bf}
	P_{\rm c} = (1-p)^n, \qquad P_{\rm b} = p(1-p)^{n-1},
\end{equation}
and the success probability of a single run is
\begin{equation}\label{dj_success_bf}
	\prsuc^{(X)} = \frac{1}{2}\left[1+(1-p)^{n-1}(1-2p)\right].
\end{equation}
Table~\ref{tab:rep_dj_bf} lists $m_{\min}$ and the corresponding threshold $K$ for $\delta=0.01$, together with Qiskit simulations of the constant ($f=0$) and balanced ($f=x_1$) oracles, performed as described in Sec.~\ref{BVbitflip}.

\begin{table}[!htbp]
	\centering
	\caption{Number of repetitions $m$ required to distinguish a constant from a balanced oracle with confidence $1-\delta=0.99$ under bit-flip noise, using the frequency of the outcome $\ket{\mathbf 0}$. $m_{\min}$ is the smallest $m$ for which the success probability of Eq.~\eqref{exact_dj} reaches $1-\delta$, and $K$ is the corresponding threshold; the analytical and simulated success probabilities at $m_{\min}$ are also listed. Each success probability is the smaller of the values for the constant ($f=0$) and the balanced ($f=x_1$) oracle. Simulated values are obtained from $2000$ trials of $m$ runs each, as described in Sec.~\ref{BVbitflip}.}
	\label{tab:rep_dj_bf}
	\begin{tabular*}{\linewidth}{c@{\extracolsep{\fill}}c cccc}
		\toprule
		$n$ & $p$ & $m_{\min}$ & $K$ & $\prsuc$ (analytical) & $\prsuc$ (Qiskit) \\
		\midrule
		10 & 0.01 & 4 & 1 & 0.9968 & 0.9980 \\
		10 & 0.05 & 11 & 2 & 0.9939 & 0.9950 \\
		20 & 0.01 & 5 & 1 & 0.9953 & 0.9950 \\
		20 & 0.05 & 20 & 2 & 0.9900 & 0.9880 \\
		30 & 0.01 & 6 & 1 & 0.9944 & 0.9935 \\
		30 & 0.05 & 36 & 2 & 0.9903 & 0.9885 \\
		
		\bottomrule
	\end{tabular*}
\end{table}

For the depolarizing channel of Eq.~\eqref{dep}, $\varepsilon=\beta$, so that
\begin{equation}\label{dj_pc_pb_dep}
	P_{\rm c} = (\alpha+\beta)^n = \left[\frac{1+(1-p)^3}{2}\right]^n, \qquad P_{\rm b} = \beta\,(\alpha+\beta)^{n-1}.
\end{equation}
Their difference is
\begin{equation}\label{m_dj_dep}
	\Delta^{\text{dep}} = P_{\rm c}-P_{\rm b} = (1-p)^3\left[\frac{1+(1-p)^3}{2}\right]^{n-1},
\end{equation}
so that the success probability of Eq.~\eqref{dj_avg} is
\begin{equation}\label{dj_success_dep}
	\prsuc^{\text{dep}} = \frac{1}{2}\left[P_{\rm c}+(1-P_{\rm b})\right] = \frac{1}{2}\left\{1+(1-p)^3\left[\frac{1+(1-p)^3}{2}\right]^{n-1}\right\},
\end{equation}
which decreases from $1$ at $p=0$ to $\tfrac{1}{2}$ at $p=1$. The results for the repeated algorithm are listed in Table~\ref{tab:rep_dj_dep}. For the phase-flip channel of Eq.~\eqref{phaseflipdef}, the same expressions hold with $\varepsilon=2p(1-p)$.

\begin{table}[!htbp]
	\centering
	\caption{As Table~\ref{tab:rep_dj_bf}, but for depolarizing noise, with $P_{\rm c}$ and $P_{\rm b}$ given by Eq.~\eqref{dj_pc_pb_dep}. Simulated values are obtained from $2000$ trials of $m$ runs each, as described in Sec.~\ref{BVbitflip}.}
	\label{tab:rep_dj_dep}
	\begin{tabular*}{\linewidth}{c@{\extracolsep{\fill}}c cccc}
		\toprule
		$n$ & $p$ & $m_{\min}$ & $K$ & $\prsuc$ (analytical) & $\prsuc$ (Qiskit) \\
		\midrule
		10 & 0.01 & 4 & 1 & 0.9904 & 0.9905 \\
		10 & 0.05 & 18 & 3 & 0.9935 & 0.9925 \\
		20 & 0.01 & 6 & 1 & 0.9946 & 0.9935 \\
		20 & 0.05 & 41 & 3 & 0.9910 & 0.9920 \\
		30 & 0.01 & 8 & 1 & 0.9956 & 0.9960 \\
		30 & 0.05 & 89 & 3 & 0.9901 & 0.9855 \\
		
		\bottomrule
	\end{tabular*}
\end{table}

\section{Results \& Discussion}\label{sec:results}

\subsection{Comparison with simulated results}\label{sec:sim}

We simulate the Bernstein--Vazirani and Deutsch--Jozsa circuits in Qiskit with the noise model described in Secs.~\ref{sec:bv} and \ref{sec:dj}. The noise channel is attached, through a noise model of the Qiskit Aer simulator, to every gate, including identity gates on the qubits on which the oracle does not act, so that each of the $n$ qubits experiences the three noise layers of Fig.~\ref{BV_noisy}. For the Bernstein--Vazirani algorithm the success probability is the probability of obtaining $\ket{\mathbf s}$, which is estimated in the simulations by the fraction of shots that return $\ket{\mathbf s}$, whereas for the Deutsch--Jozsa algorithm it is obtained from Eq.~\eqref{dj_avg}, with $P_{\rm c}$ and $P_{\rm b}$ replaced by the relative frequencies of the outcome $\ket{\mathbf 0}$ in simulations of the constant oracle $f=0$ and the balanced oracle $f(\mathbf x)=x_1$, respectively. The comparison validates the analytical model and shows how the performance of the algorithms varies with noise strength and system size.

\begin{figure}[H]
	\centering
	\begin{subfigure}[t]{0.49\textwidth}
	    \centering
        \includegraphics[width=0.96\linewidth]{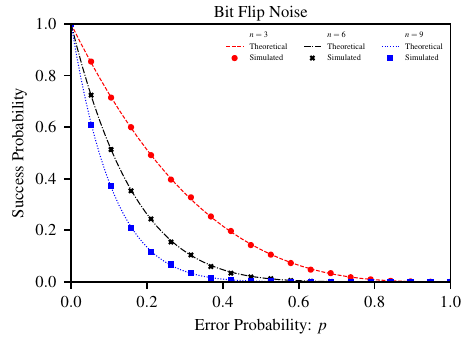}
        \includegraphics[width=0.96\linewidth]{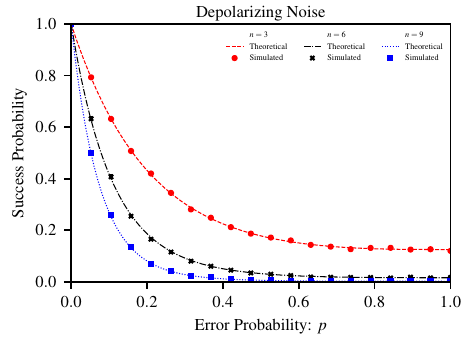}
        \caption{Bernstein--Vazirani algorithm}
	\end{subfigure}
    \medskip
    \begin{subfigure}[t]{0.49\textwidth}
	    \centering
        \includegraphics[width=0.96\linewidth]{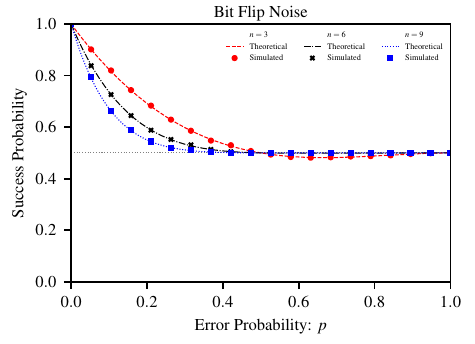}
        \includegraphics[width=0.96\linewidth]{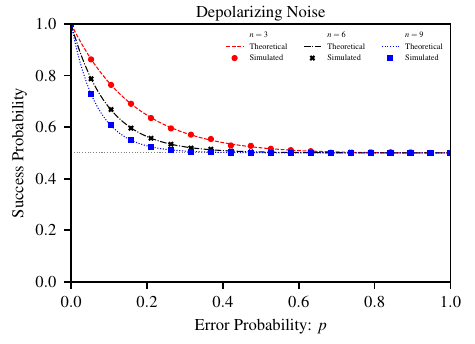}
        \caption{Deutsch--Jozsa algorithm}
	\end{subfigure}

	\caption{Comparison of theoretical and simulated success probabilities for the Bernstein--Vazirani and Deutsch--Jozsa algorithms under noise, for $n = 3$, $6$, and $9$ qubits. Within each panel, the upper and lower plots correspond to bit-flip and depolarizing noise, respectively. Theoretical predictions are shown as lines and Qiskit simulations as symbols. The simulations use shot sampling with the Qiskit Aer simulator ($10^4$ shots per circuit). In the Deutsch--Jozsa panels of this and the following figures, the dotted line marks the success probability $\tfrac{1}{2}$ of random guessing.}
	\label{pf_qiskit-vs-theory}
\end{figure}

\begin{figure}[H]
	\centering
	\begin{subfigure}[t]{0.49\textwidth}
	    \centering
        \includegraphics[width=0.96\linewidth]{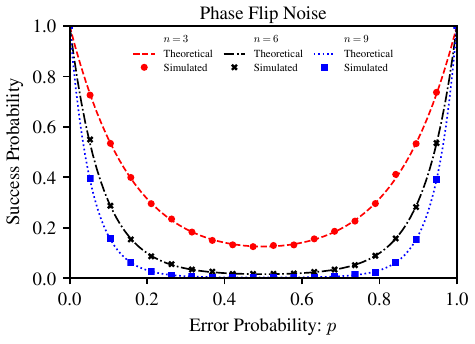}
        \includegraphics[width=0.96\linewidth]{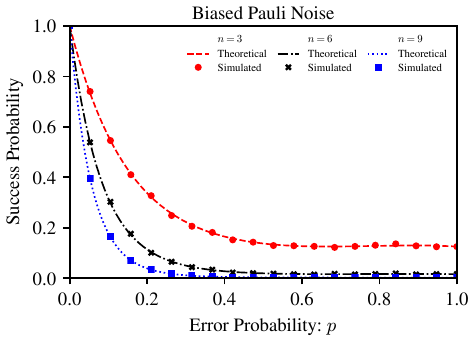}
        \caption{Bernstein--Vazirani algorithm}
	\end{subfigure}
    \medskip
    \begin{subfigure}[t]{0.49\textwidth}
	    \centering
        \includegraphics[width=0.96\linewidth]{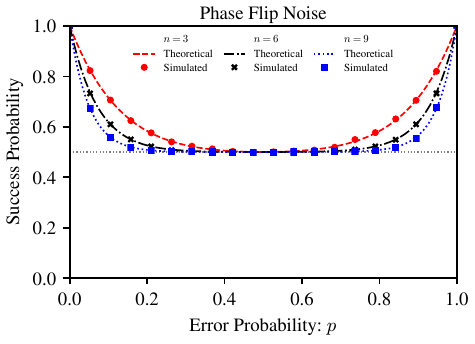}
        \includegraphics[width=0.96\linewidth]{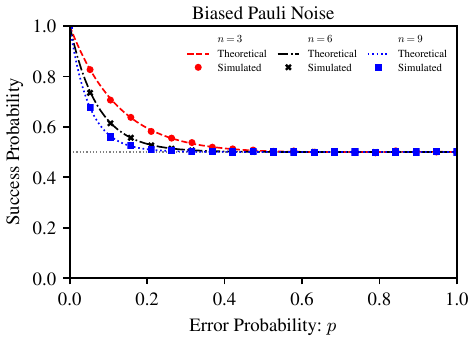}
        \caption{Deutsch--Jozsa algorithm}
	\end{subfigure}

	\caption{As Fig.~\ref{pf_qiskit-vs-theory}, but for two further Pauli channels. Within each panel, the upper plot corresponds to phase-flip noise [Eq.~\eqref{phaseflipdef}] and the lower plot to a Pauli channel biased toward dephasing, with $p_x=p_y=p/4$ and $p_z=p/2$. The lines are Eqs.~\eqref{prsuc_pauli} and \eqref{dj_success_eps} with the bit-error probability $\varepsilon$ of Eq.~\eqref{eps_def}.}
	\label{pauli_qiskit-vs-theory}
\end{figure}

Figure~\ref{pf_qiskit-vs-theory} compares the theoretical and simulated success probabilities for $n = 3$, $6$, and $9$ qubits as the error probability $p$ increases. The simulated and analytical results agree closely over the full range of $p$ for both algorithms and both noise channels. Figure~\ref{pauli_qiskit-vs-theory} shows the same comparison for two further Pauli channels. The first is the phase-flip channel of Eq.~\eqref{phaseflipdef}. The second is a channel that is biased toward dephasing, with $p_x=p_y=p/4$ and $p_z=p/2$, where $p$ is the total error probability; Eq.~\eqref{eps_def} then gives $\varepsilon=\tfrac{1}{2}\left[1-\left(1-\tfrac{3}{2}p\right)^{2}(1-p)\right]$. Under phase-flip noise the success probability is symmetric about $p=\tfrac{1}{2}$ and returns to $1$ at $p=1$, as discussed in Sec.~\ref{BVbitflip}. In both cases the simulations agree with Eqs.~\eqref{prsuc_pauli} and \eqref{dj_success_eps}, which confirms that the noise enters only through $\varepsilon$.

\subsection{Impact of noise on the algorithms for fixed qubit counts}\label{sec:noise_impact}

\begin{figure}[H]
	\centering

	\begin{subfigure}[t]{0.49\textwidth}
	    \centering
        \includegraphics[width=0.96\linewidth]{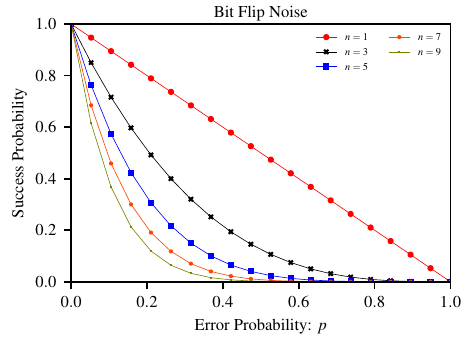}
        \includegraphics[width=0.96\linewidth]{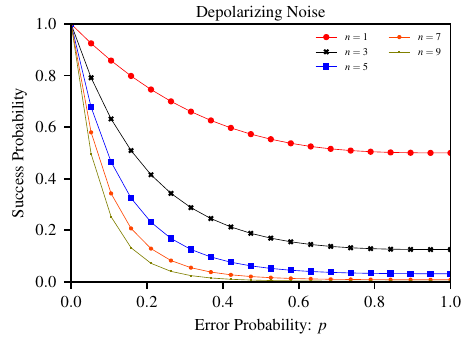}
        \caption{Bernstein--Vazirani algorithm}
	\end{subfigure}
    \medskip
    \begin{subfigure}[t]{0.49\textwidth}
	    \centering
        \includegraphics[width=0.96\linewidth]{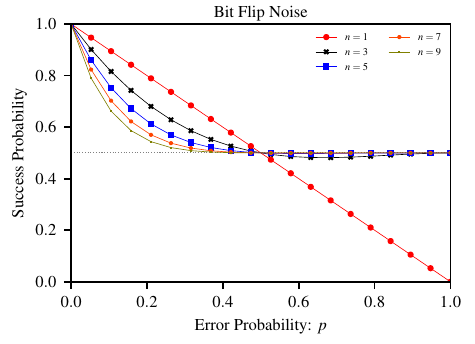}
        \includegraphics[width=0.96\linewidth]{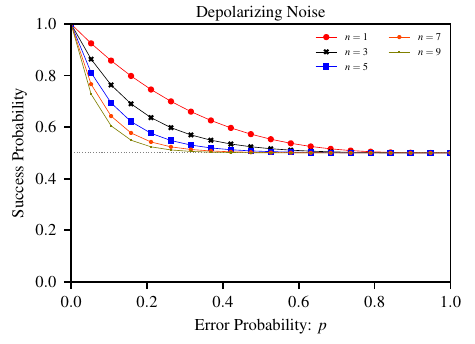}
        \caption{Deutsch--Jozsa algorithm}
	\end{subfigure}

	\caption{Success probability as a function of the error probability $p$ for several qubit counts $n$, for the Bernstein--Vazirani and Deutsch--Jozsa algorithms. Within each panel, the upper and lower plots correspond to bit-flip and depolarizing noise, respectively.}
	\label{pf_critical}
\end{figure}

To examine how noise limits performance, we plot the success probability against the error probability $p$ for several fixed qubit counts. Figures~\ref{pf_critical}--\ref{pf_fixed-p} show the analytical results in three regimes, and we summarize the quantitative conclusions they support. Figure~\ref{pf_critical} shows the success probability as a function of $p$ for several $n$.

For the Bernstein--Vazirani algorithm, let $p_{1/2}$ denote the error probability at which the success probability falls to $\tfrac{1}{2}$. Under bit-flip noise,
\begin{equation}
	\left(1-p_{1/2}\right)^n = \frac{1}{2} \quad\Longrightarrow\quad p_{1/2} = 1-2^{-1/n} \approx \frac{\ln 2}{n},
\end{equation}
and under depolarizing noise,
\begin{equation}
	\left[\frac{1+\left(1-p_{1/2}\right)^3}{2}\right]^n = \frac{1}{2} \quad\Longrightarrow\quad p_{1/2} = 1-\left(2^{1-1/n}-1\right)^{1/3} \approx \frac{2\ln 2}{3n}.
\end{equation}
For a general Pauli channel, both conditions read $\varepsilon_{1/2}=1-2^{-1/n}\approx\ln 2/n$ in terms of the bit-error probability $\varepsilon$ of Eq.~\eqref{eps_def}. On the other hand, for the Deutsch--Jozsa algorithm, the success probability of Eq.~\eqref{dj_success_eps} decreases from $1$ in the absence of noise to $\tfrac{1}{2}$, i.e., random guessing, at $\varepsilon=\tfrac{1}{2}$, which corresponds to $p=\tfrac{1}{2}$ for bit-flip and phase-flip noise and to $p=1$ for depolarizing noise. Under bit-flip noise with $p>\tfrac{1}{2}$, each output bit is more likely to be flipped than not, so that $\Delta<0$ and the success probability falls below $\tfrac{1}{2}$. To quantify the tolerable error probability, we define $\varepsilon_{3/4}$ as the value of $\varepsilon$ at which the success probability equals $\tfrac{3}{4}$, i.e., midway between the best case ($1$, no noise) and the worst case ($\tfrac{1}{2}$, random guessing). Setting $\prsuc=\tfrac{3}{4}$ in Eq.~\eqref{dj_success_eps} gives
\begin{equation}
	\left(1-\varepsilon_{3/4}\right)^{n-1}\left(1-2\varepsilon_{3/4}\right) = \frac{1}{2} \quad\Longrightarrow\quad \varepsilon_{3/4} \approx \frac{\ln 2}{n+1}.
\end{equation}
For both algorithms the tolerable error probability thus decreases as $1/n$.

\subsection{Performance in the low-noise regime}\label{sec:lownoise}

\begin{figure}[H]
	\centering

	\begin{subfigure}[t]{0.49\textwidth}
	    \centering
        \includegraphics[width=0.96\linewidth]{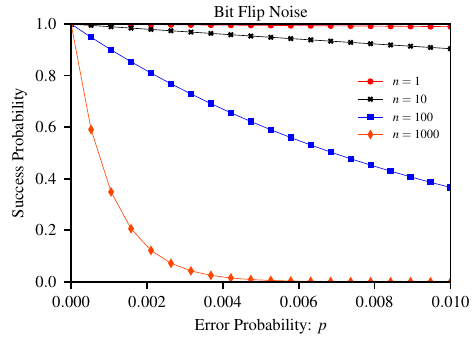}
        \includegraphics[width=0.96\linewidth]{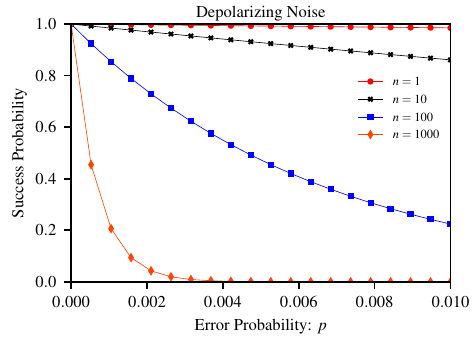}
        \caption{Bernstein--Vazirani algorithm}
	\end{subfigure}
    \medskip
    \begin{subfigure}[t]{0.49\textwidth}
	    \centering
        \includegraphics[width=0.96\linewidth]{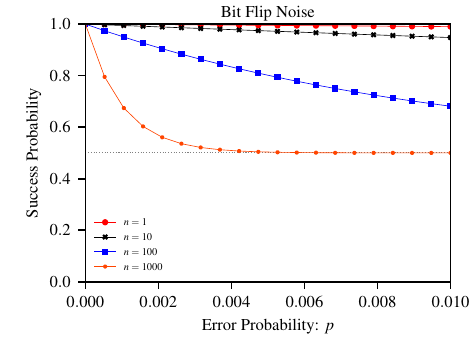}
        \includegraphics[width=0.96\linewidth]{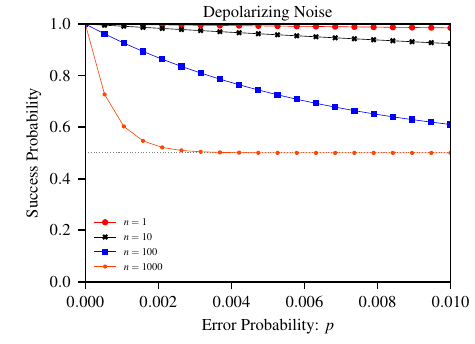}
        \caption{Deutsch--Jozsa algorithm}
	\end{subfigure}

	\caption{Success probability versus the error probability $p$ in the low-noise regime $0 \leq p \leq 0.01$, for several qubit counts $n$ and both algorithms. Within each panel, the upper and lower plots correspond to bit-flip and depolarizing noise, respectively.}
	\label{pf_low-noise}
\end{figure}

Figure~\ref{pf_low-noise} shows the success probability across the low-noise regime $0 \leq p \leq 0.01$ for several qubit counts $n$. Whenever $np\ll1$, the success probability decreases linearly with $p$. As $n$ increases, however, the decrease is no longer linear even for small error probabilities, as seen in Fig.~\ref{pf_low-noise} for $n=100$ and $n=1000$.

\subsection{Impact of system size at fixed noise}\label{sec:size}

\begin{figure}[H]
	\centering

	\begin{subfigure}[t]{0.49\textwidth}
	    \centering
        \includegraphics[width=0.96\linewidth]{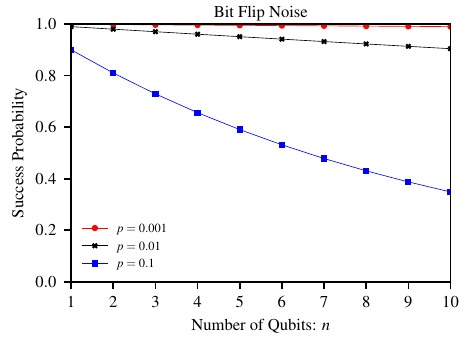}
        \includegraphics[width=0.96\linewidth]{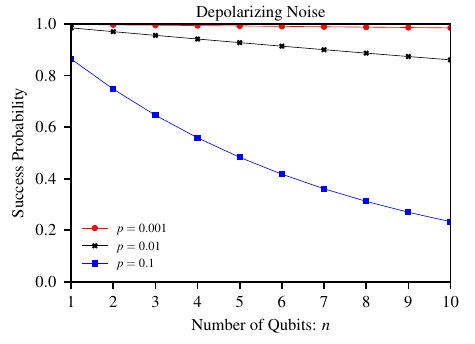}
        \caption{Bernstein--Vazirani algorithm}
	\end{subfigure}
    \medskip
    \begin{subfigure}[t]{0.49\textwidth}
	    \centering
        \includegraphics[width=0.96\linewidth]{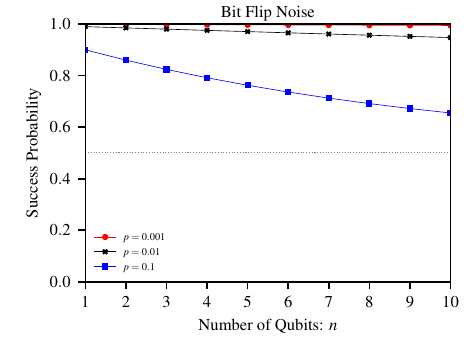}
        \includegraphics[width=0.96\linewidth]{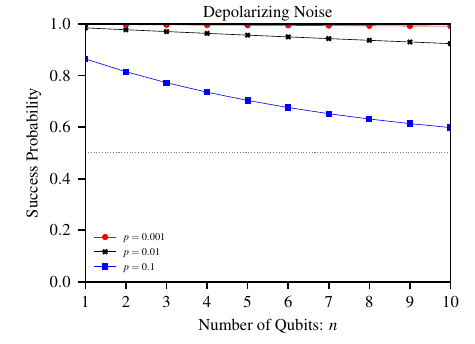}
        \caption{Deutsch--Jozsa algorithm}
	\end{subfigure}

	\caption{Success probability as a function of the number of qubits $n$ for fixed error probabilities $p = 0.001$, $0.01$, and $0.1$, for both algorithms. Within each panel, the upper and lower plots correspond to bit-flip and depolarizing noise, respectively.}
	\label{pf_fixed-p}
\end{figure}

To probe scalability, we fix the error probability $p$ and follow the success probability as the number of qubits $n$ grows. Figure~\ref{pf_fixed-p} shows this behavior for representative noise levels $p = 0.001$, $0.01$, and $0.1$.

At fixed $p$ the Bernstein--Vazirani success probability decays exponentially with $n$. Under bit-flip noise, a success probability of at least $P_0$ requires
\begin{equation}\label{eq:nmax_bf}
	(1-p)^n \geq P_0 \quad\Longrightarrow\quad n \leq \frac{\ln(1/P_0)}{\ln[1/(1-p)]} \approx \frac{\ln(1/P_0)}{p},
\end{equation}
and under depolarizing noise,
\begin{equation}\label{eq:nmax_dep}
	(\alpha+\beta)^n \geq P_0 \quad\Longrightarrow\quad n \leq \frac{\ln(1/P_0)}{\ln[1/(\alpha+\beta)]} \approx \frac{2\ln(1/P_0)}{3p},
\end{equation}
where the approximations hold for small $p$. For $P_0=0.9$, these give $n\leq105$ (bit flip) and $n\leq70$ (depolarizing) at $p=0.001$, and $n\leq10$ and $n\leq7$ at $p=0.01$. Equivalently, the error probability must decrease as $1/n$ to maintain a fixed accuracy. The Deutsch--Jozsa success probability of Eq.~\eqref{dj_success_eps} also decreases with $n$ and approaches $\tfrac{1}{2}$; a success probability of at least $0.9$ requires $n\leq222$ (bit flip) and $n\leq147$ (depolarizing) at $p=0.001$, and $n\leq21$ and $n\leq13$ at $p=0.01$.

\subsection{Repetitions and comparison with classical algorithms}\label{sec:repetitions}
In practice, a noisy circuit is repeated and its outcomes are post-processed, so the relevant resource is the total number of runs, each of which requires a single oracle query. Equations~\eqref{exact_bv} and \eqref{exact_dj} give the success probability after $m$ runs, and hence the minimum number of runs $m_{\min}$ for a required confidence $1-\delta$, for both algorithms and for any Pauli channel. Tables~\ref{tab:repetitions}--\ref{tab:rep_dj_dep} list $m_{\min}$ for bit-flip and depolarizing noise and compare the analytical success probabilities with Qiskit simulations; in all cases the two agree to within about two standard errors of the simulations. 

The tables also reveal a qualitative difference between the two algorithms. For the Bernstein--Vazirani algorithm the answer is encoded in $n$ output bits that can be decoded independently, and the required number of runs grows only slowly with $n$. For example, at $p=0.3$, $m_{\min}$ increases only from $55$ to $67$ under bit-flip noise and from $77$ to $93$ under depolarizing noise as $n$ increases from $10$ to $30$.

\begin{figure}[!htbp]
	\centering
	\includegraphics[width=0.8\linewidth]{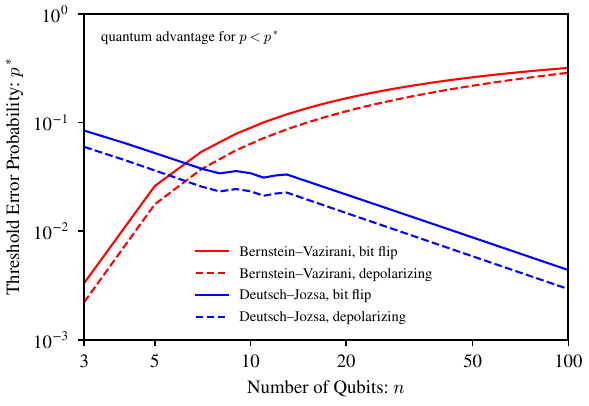}
	\caption{Threshold error probability $p^{*}$ below which the noisy quantum algorithm beats its classical counterpart, as a function of the number of qubits $n$, for a confidence $1-\delta=0.99$. Bernstein--Vazirani algorithm (red): $m_{\min}<n$, shown for odd $n$. Deutsch--Jozsa algorithm (blue): $m_{\min}<8$, the number of queries of the randomized classical algorithm. Solid and dashed lines correspond to bit-flip and depolarizing noise, respectively. The threshold of the Bernstein--Vazirani algorithm increases with $n$, whereas that of the Deutsch--Jozsa algorithm decreases as $1/n$; the small steps in the latter near $n\approx10$ arise from changes of the optimal number of runs and threshold $K$.}
	\label{fig:advantage}
\end{figure}

A classical algorithm requires $n$ queries, one per bit, so that the noisy quantum algorithm has a query advantage only if the required number of runs is smaller than $n$. This is not the case in the example above, but since $m_{\min}$ grows slowly with $n$ whereas the classical cost grows linearly, the advantage is recovered at smaller $p$ or larger $n$. We denote by $p^{*}(n)$ the threshold error probability below which $m_{\min}<n$, i.e., below which the noisy algorithm retains its query advantage. Figure~\ref{fig:advantage} shows $p^{*}$ for $\delta=0.01$. Under bit-flip noise it increases with $n$, from about $0.1$ at $n=11$ to about $0.3$ at $n=101$, so that the range of error probabilities over which the advantage is retained widens as the system size increases. This is consistent with the logarithmic growth of the number of quantum queries that was found for the learning of parity functions from noisy examples~\cite{cross2015,riste2017}. Here the comparison is with the noiseless classical algorithm, and the threshold is given explicitly.

For the Deutsch--Jozsa algorithm, in contrast, the verdict rests on the single outcome $\ket{\mathbf 0}$, whose probability decays exponentially with $n$. Consequently, $P_{\rm c}$ and $P_{\rm b}$ become increasingly difficult to distinguish, and the required number of runs grows rapidly with $n$. For example, at $p=0.05$, $m_{\min}$ increases from $11$ to $36$ under bit-flip noise and from $18$ to $89$ under depolarizing noise as $n$ increases from $10$ to $30$.

Classically, the Deutsch--Jozsa problem can be solved with a small error probability using only a few queries~\cite{neilson}. One evaluates $f$ on $k$ randomly chosen inputs and declares $f$ balanced if two of the outputs differ, and constant if all outputs agree. This verdict is wrong only if $f$ is balanced and all $k$ outputs happen to coincide, which occurs with probability
\begin{equation}\label{eq:classical_dj}
	\Pr(\text{error}) = 2\left(\frac{1}{2}\right)^{k} = 2^{1-k},
\end{equation}
independently of $n$. An error probability of $\delta=0.01$ therefore requires only $k=8$ classical queries, and the noisy quantum algorithm retains a query advantage only if $m_{\min}<8$. The corresponding threshold $p^{*}(n)$ is also shown in Fig.~\ref{fig:advantage}. For $13\leq n\leq100$ it decreases with $n$ as $p^{*}\approx0.44/n$ under bit-flip noise and $p^{*}\approx0.29/n$ under depolarizing noise. Outside this range the optimal number of runs and the optimal threshold $K$ of Eq.~\eqref{exact_dj} change, and with them the coefficient, but for all $n\geq3$ the threshold lies between $0.25/n$ and $0.73/n$ for bit-flip noise. The small steps near $n\approx10$ in Fig.~\ref{fig:advantage} have the following origin. The condition for a constant oracle bounds the product $n\varepsilon$, whereas the condition for a balanced oracle bounds $P_{\rm b}$ and hence $\varepsilon$ itself. A smaller threshold $K$ tolerates a larger $n\varepsilon$, but it is admissible only once $\varepsilon$ is small enough to satisfy the second condition, i.e., for sufficiently large $n$. As $n$ increases, the optimal pair $(m,K)$ therefore changes from $(7,2)$ to $(6,1)$ at $n=8$ and to $(7,1)$ at $n=12$, and where the condition for the balanced oracle is the binding one, $p^{*}$ is nearly independent of $n$. Since $f(\mathbf x)=x_1$ is the least favorable balanced oracle, this threshold applies to every balanced oracle. For both algorithms the thresholds depend on the channel only through $\varepsilon$, so that $p^{*}$ for any other Pauli channel follows from the bit-flip curves of Fig.~\ref{fig:advantage}, which give $\varepsilon^{*}$, and Eq.~\eqref{eps_def}.

These opposite trends have a simple origin. A single Bernstein--Vazirani run returns all $n$ bits of $\mathbf s$, so that its cost grows much more slowly with $n$ than the $n$ classical queries, whereas the Deutsch--Jozsa verdict requires all $n$ qubits to be free of errors while its classical cost stays fixed.

The $1/n$ dependence is a property of the standard readout of the Deutsch--Jozsa algorithm, which tests for the single outcome $\ket{\mathbf 0}$ and makes no assumption about the balanced function. It is not a property of the circuit. For $f(\mathbf x)=x_1$ the circuit coincides with the Bernstein--Vazirani circuit for $\mathbf s=(1,0,\dots,0)$, so that the output could instead be decoded by the bitwise majority vote of Sec.~\ref{sec:bv_pauli}, and the oracle declared constant if the decoded string is $\mathbf 0$. The verdict is then correct with a probability of at least $\left[P_{\rm bit}(m)\right]^n$ for either type of oracle, and seven runs suffice for $\varepsilon<\varepsilon^{*}\approx0.13\,n^{-1/4}$. At $n=100$ this gives $\varepsilon^{*}\approx0.042$, compared with $\varepsilon^{*}\approx0.0044$ for the standard readout. This readout, however, relies on the ideal output being a single string. It fails for balanced functions whose ideal output is spread over many strings. For the majority function of an odd number $n\geq5$ of bits, for example, each ideal output bit equals $1$ with a probability smaller than $\tfrac{1}{2}$, so that the vote returns $\mathbf 0$, and the oracle is wrongly declared constant, with a probability that approaches $1$ as the number of runs increases. The thresholds of Fig.~\ref{fig:advantage} therefore refer to the standard readout, which is valid for every balanced function.

The noisy Deutsch--Jozsa algorithm with its standard readout thus loses its query advantage once the error probability exceeds a value that decreases as $1/n$, whereas the Bernstein--Vazirani algorithm retains its advantage over a wide range of error probabilities. For near-term devices, this contrast indicates that algorithms whose answer is encoded in many independently decodable output bits are considerably more robust to noise than algorithms whose verdict depends on a single global outcome.

\section{Conclusions and Discussion}\label{sec:conclusions}

Using the density-matrix formalism, we derived exact closed-form expressions for the success probability of the Bernstein--Vazirani and Deutsch--Jozsa algorithms under a general single-qubit Pauli channel acting on each qubit after every gate, with bit-flip, phase-flip, and depolarizing noise as special cases, as functions of the single-qubit error probabilities and the number of qubits $n$, and validated them against Qiskit simulations of four Pauli channels, finding agreement within the statistical uncertainty. The noise enters only through the probability $\varepsilon$ with which each output bit is flipped. The channels differ qualitatively: under bit-flip noise the most probable outcome is the hidden string for $p<\tfrac{1}{2}$ and its complement for $p>\tfrac{1}{2}$, from which the hidden string is recovered by classical post-processing, so that the algorithm fails only at $p=\tfrac{1}{2}$, whereas the depolarizing and phase-flip channels never bias the output toward a wrong string. For the Deutsch--Jozsa algorithm, the closed-form expressions are given for the single-variable oracle $f(\mathbf x)=x_1$, which is the least robust of all balanced oracles for $\varepsilon<\tfrac{1}{2}$, as follows from the result for an arbitrary balanced function and as Qiskit simulations of other balanced oracles, including an entangling one, confirm. Current devices operate far below $p=\tfrac{1}{2}$, so the practically relevant results are the tolerable error probability and the number of repetitions. For a single run, the tolerable error probability of both algorithms decreases as $1/n$. With repetitions, the threshold error probability below which the Bernstein--Vazirani algorithm retains its query advantage increases with $n$, whereas that of the Deutsch--Jozsa algorithm with its standard readout, which tests for the single outcome $\ket{\mathbf 0}$, decreases as $1/n$. Qubit-dependent error probabilities change the single-run success probabilities only at second order in their spread, so that $p$ may be read as the mean error probability (Appendix~\ref{app:nonuniform}). These conclusions apply specifically to the assumed independent Pauli channels; extending the analysis to spatially or temporally correlated and non-Markovian noise, and connecting these expressions to error-mitigation strategies, are natural directions for future work. More broadly, our results show that the robustness of a quantum algorithm to noise depends on how its answer is read out: when the answer is encoded in many independently decodable output bits, as in the Bernstein--Vazirani algorithm, the threshold error probability for retaining the query advantage increases with the system size, whereas a verdict that depends on a single global outcome, as in the standard readout of the Deutsch--Jozsa algorithm, requires an error probability that decreases as $1/n$.

\backmatter

\begin{appendices}
\providecommand{\theHequation}{}\renewcommand{\theHequation}{app.\thesection.\arabic{equation}}

\section{Derivation of the success probabilities}\label{app:deriv}
\setcounter{equation}{0}
This appendix contains the derivations of the single-run results quoted in Secs.~\ref{sec:bv_noise} and \ref{sec:dj_noise}, for the Pauli channel of Eq.~\eqref{paulidef}.

\subsection{Bernstein--Vazirani algorithm}\label{app:deriv_bv}
It is convenient to write a single-qubit state in the Pauli basis as $\rho=\tfrac{1}{2}\left(\mathbb{I}+\mathbf n\cdot\boldsymbol\sigma\right)$, where $\mathbf n=(n_x,n_y,n_z)$ is the Bloch vector and $\boldsymbol\sigma=(X,Y,Z)$ is the vector of Pauli matrices. Since each Pauli operator commutes with itself and anticommutes with the other two, Eq.~\eqref{paulidef} becomes
\begin{equation}\label{pauli_bloch}
	\mathcal{E}(\rho) = \frac{1}{2}\left(\mathbb{I} + r_x n_x X + r_y n_y Y + r_z n_z Z\right),
\end{equation}
with
\begin{equation}\label{r_def}
	r_x = 1-2(p_y+p_z), \qquad r_y = 1-2(p_x+p_z), \qquad r_z = 1-2(p_x+p_y),
\end{equation}
i.e., the channel scales the three components of the Bloch vector by $r_x$, $r_y$, and $r_z$. With this channel acting on each qubit after every gate (Fig.~\ref{BV_noisy}), the single-qubit state evolves as
\begin{align}
	\phi_1^i & = H\op{0}{0}H = \tfrac{1}{2}\left(\mathbb{I}+X\right) \\
	\phi_2^i & = \mathcal{E}(\phi_1^i) = \tfrac{1}{2}\left(\mathbb{I}+r_x X\right) \\
	\phi_3^i & = Z^{s_i} \phi_2^i Z^{s_i} = \tfrac{1}{2}\left[\mathbb{I}+r_x(-1)^{s_i}X\right] \\
	\phi_4^i & = \mathcal{E}(\phi_3^i) = \tfrac{1}{2}\left[\mathbb{I}+r_x^{2}(-1)^{s_i}X\right] \\
	\phi_5^i & = H\phi_4^i H = \tfrac{1}{2}\left[\mathbb{I}+r_x^{2}(-1)^{s_i}Z\right] \\
	\phi_6^i & = \mathcal{E}(\phi_5^i) = \tfrac{1}{2}\left[\mathbb{I}+r_x^{2}r_z(-1)^{s_i}Z\right] = (1-\varepsilon)\op{s_i}{s_i} + \varepsilon \op{\bar{s_i}}{\bar{s_i}} \label{phi6_single}
\end{align}%
where  $\bar{s_i}$ denotes the complement of $s_i$, $\varepsilon=\tfrac{1}{2}\left(1-r_x^{2}r_z\right)$ is the bit-error probability of Eq.~\eqref{eps_def}, and in the last step we used $\mathbb{I}=\op{s_i}{s_i}+\op{\bar{s_i}}{\bar{s_i}}$ and $(-1)^{s_i}Z=\op{s_i}{s_i}-\op{\bar{s_i}}{\bar{s_i}}$. The first two noise layers act on a state that has only an $X$ component and scale it by $r_x$ each, and the final layer, acting after the Hadamard, scales the $Z$ component by $r_z$.
For the full $n$-qubit circuit, the pre-measurement state is
\begin{equation}\label{phi6_bf}
	\phi_6 = \bigotimes_{i=1}^n \left[(1-\varepsilon)\,\op{s_i}{s_i} + \varepsilon\op{\bar{s_i}}{\bar{s_i}}\right].
\end{equation}
The success probability is the overlap of $\phi_6$ with the target $\ket{\mathbf s}$. Using  $\braket{s_i}{\bar{s_i}} = 0$, this overlap factorizes into identical single-qubit contributions:
\begin{equation}\label{app_prsuc_bf}
	\begin{aligned}
		\prsuc & = \mel{\mathbf s}{\phi_6}{\mathbf s} = \prod_{i=1}^n (1-\varepsilon) = (1-\varepsilon)^n.
	\end{aligned}
\end{equation}
This is the result quoted in Eq.~\eqref{prsuc_pauli}. The special cases follow from Eq.~\eqref{r_def}. For bit-flip noise ($p_x=p$), $r_x=1$ and $r_z=1-2p$: the first two noise layers leave the state unchanged, only the final layer alters it, and $\varepsilon=p$ [Eq.~\eqref{prsuc_bf}]. For phase-flip noise ($p_z=p$), $r_x=1-2p$ and $r_z=1$: only the first two layers alter the state, and $\varepsilon=\tfrac{1}{2}\left[1-(1-2p)^2\right]=2p(1-p)$ [Eq.~\eqref{prsuc_pf}]. For depolarizing noise ($p_x=p_y=p_z=p/4$), $r_x=r_y=r_z=1-p$: each layer leaves the identity component untouched and rescales the traceless part by $1-p$, so that its weight accumulates as the third power $(1-p)^3$, and $\varepsilon=\tfrac{1}{2}\left[1-(1-p)^3\right]=\beta$ [Eq.~\eqref{prsuc_dep}].

\subsection{Deutsch--Jozsa algorithm}\label{app:deriv_dj}
Here we derive the probability $\Pr(\ket{\mathbf 0})$ of obtaining $\ket{\mathbf 0}$ for an arbitrary Boolean function $f$, and from it the probability $P_{\rm b}$ for a balanced oracle. Since the oracle may be entangling, the state is in general not a product state, and we follow the action of the channel on the operators $\op{x}{y}$ instead of on single-qubit Bloch vectors. For a single qubit, with $x_i,y_i\in\{0,1\}$, $u_i=x_i\oplus y_i$, and $p_0 = 1-p_x-p_y-p_z$, Eq.~\eqref{paulidef} gives
\begin{equation}\label{pauli_unit}
	\mathcal{E}\left(\op{x_i}{y_i}\right) = \left[p_0+(-1)^{u_i}p_z\right]\op{x_i}{y_i} + \left[p_x+(-1)^{u_i}p_y\right]\op{\bar x_i}{\bar y_i},
\end{equation}
where a bar denotes the complement. The $Z$ component of the noise thus changes the sign of the coherences, whereas the $X$ and $Y$ components flip both indices.

After the first Hadamard layer the state is $\rho_1=\op{+}{+}^{\otimes n}$, with each qubit along the $x$ axis of the Bloch sphere. By Eq.~\eqref{pauli_bloch} the first noise layer scales each Bloch vector by $r_x$, and the oracle then multiplies $\op{x}{y}$ by $(-1)^{f(x)+f(y)}$:
\begin{align}
    \rho_1^\prime & = \mathcal E(\rho_1) = \bigotimes_{i=1}^{n} \frac{1}{2} (\mathbb{I} + r_x X)_i = \frac{1}{N} \sum_{x,y} r_x^{d(x,y)} \op{x}{y} \\
    \rho_2 & = U_f\,\rho_1^\prime\,U_f = \frac{1}{N} \sum_{x,y} r_x^{d(x,y)} (-1)^{f(x) + f(y)} \op{x}{y} 
\end{align}
where the sums run over $\mathbf x,\mathbf y\in\{0,1\}^n$ and $d(x,y)$ is the Hamming distance between $\mathbf x$ and $\mathbf y$. For the second noise layer we apply Eq.~\eqref{pauli_unit} to every qubit. Let $e_i=1$ if the indices of the $i$\textsuperscript{th} qubit are flipped and $e_i=0$ otherwise. Then
\begin{equation}\label{rho2prime}
    \rho_2^\prime = \mathcal E(\rho_2) = \frac{1}{N} \sum_{\mathbf e}\sum_{x,y} r_x^{d(x,y)}\, W_{\mathbf e}(\mathbf u)\, (-1)^{f(x \oplus e) + f(y\oplus e)} \op{x}{y}, \qquad W_{\mathbf e}(\mathbf u)=\prod_{i=1}^{n} w_{e_i}(u_i),
\end{equation}
with $\mathbf u=\mathbf x\oplus\mathbf y$, $w_0(u_i)=p_0+(-1)^{u_i}p_z$, and $w_1(u_i)=p_x+(-1)^{u_i}p_y$. Here we have relabeled $\mathbf x\to\mathbf x\oplus\mathbf e$ and $\mathbf y\to\mathbf y\oplus\mathbf e$ in the sums, which leaves $\mathbf u$ and $d(x,y)$ unchanged. The bit-flip component of the noise thus shifts the argument of $f$. After the final Hadamard layer the state is $\rho_3=H^{\otimes n}\rho_2^\prime H^{\otimes n}$, with diagonal elements
\begin{equation}
    \left[\rho_3\right]_{a,a} = \frac{1}{N^2}\sum_{\mathbf e}\sum_{x,y} r_x^{d(x,y)}\, W_{\mathbf e}(\mathbf u)\, (-1)^{f(x \oplus e) + f(y\oplus e)}\,(-1)^{\mathbf a\cdot\mathbf u}.
\end{equation}
The third noise layer acts qubit-wise, $\rho_3^\prime=\mathcal E(\rho_3)$. The probability that $\rho_3'$ is found in $\ket{0}^{\otimes n}$ is
\begin{align}
    \Pr(\ket{\mathbf 0}) & = \sum_{a,b}\left[\rho_3\right]_{a,b} K(a,b), \\
    K(a,b) & = \mel{0^{\otimes n}}{\mathcal E(\op{a}{b})}{0^{\otimes n}} = \prod_{i=1}^{n}\mel{0}{\mathcal E(\op{a_i}{b_i})}{0}.
\end{align}
By Eq.~\eqref{pauli_unit}, $K(a,b)$ vanishes unless $a_i=b_i~\forall~i$, so that only the diagonal entries of $\rho_3$ feed the $\ket{0}^{\otimes n}$ population, and
\begin{align}
    K(a,a) & = \left(\frac{1+r_z}{2} \right)^{n - w(a)} \left( \frac{1-r_z}{2} \right)^{w(a)}, 
\end{align}
where $w(a)$ is the Hamming weight of $\mathbf a$. Using
\begin{equation}
	\sum_a K(a,a)\,(-1)^{\mathbf a\cdot\mathbf u} = \prod_{i=1}^{n}\left[\frac{1+r_z}{2}+(-1)^{u_i}\frac{1-r_z}{2}\right] = r_z^{d(x,y)},
\end{equation}
we obtain
\begin{align}
    \Pr(\ket {\mathbf 0}) & = \sum_a K(a,a) \left[\rho_3\right]_{a,a} = \frac{1}{N^2}\sum_{\mathbf e}\sum_{x,y} (r_x r_z)^{d(x,y)}\, W_{\mathbf e}(\mathbf u)\, (-1)^{f(x \oplus e) + f(y\oplus e)}  \nonumber\\
    & = \frac{1}{N^2}\sum_{x,y} (r_x r_z)^{d(x,y)}\, (-1)^{f(x) + f(y)} \sum_{\mathbf e} W_{\mathbf e}(\mathbf u) \nonumber\\
    & = \frac{1}{N^2} \sum_{x,y} \left(r_x^{2} r_z\right)^{d(x,y)} (-1)^{f(x) + f(y)}. \label{dj_sum}
\end{align}
In the second line the relabeling of $\mathbf x$ and $\mathbf y$ has been reversed, which removes the shift from the argument of $f$, and in the third we used $\sum_{\mathbf e}W_{\mathbf e}(\mathbf u)=\prod_i\left[w_0(u_i)+w_1(u_i)\right]=r_x^{d(x,y)}$. Equation~\eqref{dj_sum} holds for every Boolean function $f$, whether or not its oracle is entangling. For bit-flip noise $r_x^2r_z=1-2p$, and for depolarizing noise $r_x^2r_z=(1-p)^3$.

Equation~\eqref{dj_sum} can be written in terms of the ideal output distribution. In the absence of noise the outcome $\mathbf a$ is obtained with probability $\lvert\hat f(\mathbf a)\rvert^2$, where
\begin{equation}
	\hat f(\mathbf a) = \frac{1}{N}\sum_{x}(-1)^{f(x)+\mathbf a\cdot\mathbf x}
\end{equation}
and $\sum_a\lvert\hat f(\mathbf a)\rvert^2=1$ [cf.\ Eq.~\eqref{dj_rho3}]. With $r_x^{2}r_z=1-2\varepsilon$ and
\begin{equation}
	(1-2\varepsilon)^{d(x,y)} = \prod_{i=1}^{n}\left[(1-\varepsilon)+(-1)^{u_i}\varepsilon\right] = \sum_a \varepsilon^{w(a)}(1-\varepsilon)^{n-w(a)}\,(-1)^{\mathbf a\cdot\mathbf u},
\end{equation}
Eq.~\eqref{dj_sum} becomes
\begin{equation}\label{dj_spectrum}
	\Pr(\ket{\mathbf 0}) = \sum_a \varepsilon^{w(a)}(1-\varepsilon)^{n-w(a)}\,\lvert\hat f(\mathbf a)\rvert^2 .
\end{equation}
The ideal outcome $\mathbf a$ is thus converted into $\mathbf 0$ if exactly its $w(a)$ nonzero bits are flipped, each with probability $\varepsilon$. For a constant function $\lvert\hat f(\mathbf 0)\rvert=1$, and Eq.~\eqref{dj_spectrum} gives $P_{\rm c}=(1-\varepsilon)^n$. For a balanced function $\hat f(\mathbf 0)=0$, so that only strings with $w(a)\geq1$ contribute. Since $\varepsilon^{w}(1-\varepsilon)^{n-w}$ decreases with $w$ for $\varepsilon<\tfrac{1}{2}$,
\begin{equation}\label{dj_bound}
	P_{\rm b} = \sum_{a\neq 0} \varepsilon^{w(a)}(1-\varepsilon)^{n-w(a)}\,\lvert\hat f(\mathbf a)\rvert^2 \;\leq\; \varepsilon(1-\varepsilon)^{n-1}
\end{equation}
for every balanced function, with equality if the ideal output has Hamming weight one. For the oracle $f(\mathbf x)=x_1$ considered in the main text, the ideal output is $\mathbf a=(1,0,\dots,0)$, and
\begin{equation}\label{dj_bf}
	P_{\rm b} = \varepsilon(1-\varepsilon)^{n-1}.
\end{equation}
This is the expression for $P_{\rm b}$ in Eq.~\eqref{dj_pc_pb}; it reduces to $p(1-p)^{n-1}$ for bit-flip noise and to $\beta(1-\beta)^{n-1}$ for depolarizing noise [Eqs.~\eqref{dj_pc_pb_bf} and \eqref{dj_pc_pb_dep}]. For $f(\mathbf x)=x_1\oplus x_2$ the ideal output is $(1,1,0,\dots,0)$, so that $P_{\rm b}=\varepsilon^{2}(1-\varepsilon)^{n-2}$. For the entangling oracle $f(\mathbf x)=x_1\oplus x_2x_3$ the ideal output is one of the four strings $(1,a_2,a_3,0,\dots,0)$, each with probability $\tfrac{1}{4}$, so that $P_{\rm b}=\tfrac{1}{4}\varepsilon(1-\varepsilon)^{n-3}$. These are the expressions used in Fig.~\ref{fig:oracle_comparison}.

\section{Qubit-dependent error probabilities}\label{app:nonuniform}
\subsection{Bernstein--Vazirani algorithm}\label{app:nonuniform_bv}
\setcounter{equation}{0}
In the main text every qubit has the same error probability $p$. Here we show that qubit-dependent error probabilities $p_i$ change the success probability only at second order in their spread. The success probability of the Bernstein--Vazirani algorithm is a product of single-qubit factors $g=1-\varepsilon$; for bit-flip and depolarizing noise,
\begin{equation}\label{app_prod}
	\prsuc = \prod_{i=1}^{n} g(p_i), \qquad g^{(X)}(p) = 1-p, \qquad g^{\rm dep}(p) = \alpha+\beta = \frac{1+(1-p)^3}{2},
\end{equation}
since each qubit evolves independently (Sec.~\ref{sec:bv_noise}). We write $p_i=\bar p+\delta_i$, where $\bar p=\frac{1}{n}\sum_i p_i$ is the mean error probability, so that $\sum_i\delta_i=0$, and define the variance $\sigma^2=\frac{1}{n}\sum_i\delta_i^2$. Expanding $\ln g(p_i)$ about $\bar p$,
\begin{equation}
	\ln g(p_i) = \ln g(\bar p) + c_1\,\delta_i + \tfrac{1}{2}\,c_2\,\delta_i^2 + O(\delta_i^3), \qquad c_k = \frac{d^k \ln g}{dp^k}\Big|_{p=\bar p},
\end{equation}
and summing over $i$, the first-order terms cancel because $\sum_i\delta_i=0$. Hence
\begin{equation}\label{app_nonuniform}
	\prsuc = g(\bar p)^{\,n}\left[1+\tfrac{1}{2}\,c_2\,n\,\sigma^2+O(n\,\delta^3)\right],
\end{equation}
where $g(\bar p)^n$ is the uniform result of Eqs.~\eqref{prsuc_bf} and \eqref{prsuc_dep} evaluated at the mean error probability, and
\begin{equation}
	c_2 = \begin{cases}
		-1/(1-\bar p)^2 & \text{for bit-flip noise,}\\
		\tfrac{3}{4}+O(\bar p) & \text{for depolarizing noise.}
	\end{cases}
\end{equation}
The spread thus enters only at second order, and its effect is the same as that of a small shift of the error probability. For small $\bar p$, the device behaves as one with uniform error probability
\begin{equation}\label{app_peff}
	p_{\rm eff} \approx \begin{cases}
		\bar p + \tfrac{1}{2}\sigma^2 & \text{for bit-flip noise,}\\
		\bar p - \tfrac{1}{4}\sigma^2 & \text{for depolarizing noise,}
	\end{cases}
\end{equation}
independently of $n$. If the spread is not larger than the mean, $\sigma\lesssim\bar p$, as is typically the case, the shift is much smaller than $\bar p$ itself, and the mean $\bar p$ can be taken as the effective error probability of the device. The results of the main text therefore apply to devices with qubit-dependent error probabilities, with $p$ read as the mean error probability $\bar p$.

\subsection{Deutsch--Jozsa algorithm}\label{app:nonuniform_dj}
For the oracle $f(\mathbf x)=x_1$ the circuit coincides with the Bernstein--Vazirani circuit for $\mathbf s=(1,0,\dots,0)$ (Sec.~\ref{sec:bv_noise}), so that each qubit evolves independently. The $i$\textsuperscript{th} measured bit is therefore wrong with its own probability
\begin{equation}\label{app_eps}
	\varepsilon_i = \begin{cases}
		p_i & \text{for bit-flip noise,}\\
		\beta(p_i)=\tfrac{1}{2}\left[1-(1-p_i)^3\right] & \text{for depolarizing noise,}
	\end{cases}
\end{equation}
and correct with probability $1-\varepsilon_i$.

For a constant oracle the correct outcome is $\ket{\mathbf 0}$, which is obtained only if no bit is wrong:
\begin{equation}
	\Pr(\ket{\mathbf 0}) = \prod_{i=1}^{n}(1-\varepsilon_i).
\end{equation}
This has the form of Eq.~\eqref{app_prod}. As for the Bernstein--Vazirani algorithm, it therefore equals the uniform result evaluated at the mean error probability $\bar p$, up to the small shift of Eq.~\eqref{app_peff}.

For the balanced oracle $f(\mathbf x)=x_1$ the noiseless outcome is $\ket{10\cdots0}$. The incorrect outcome $\ket{\mathbf 0}$ is obtained only if the first bit is wrong and all other bits are correct:
\begin{equation}\label{app_dj_nonuniform}
	\Pr(\ket{\mathbf 0}) = \varepsilon_1\prod_{i=2}^{n}(1-\varepsilon_i).
\end{equation}
The two factors behave differently. The product over the remaining $n-1$ qubits again has the form of Eq.~\eqref{app_prod} and depends only on their mean error probability $\bar p$. The prefactor $\varepsilon_1$, however, belongs to the first qubit alone, since this is the only qubit that carries the information about $f$. The expressions for $P_{\rm b}$ in Eqs.~\eqref{dj_pc_pb_bf} and \eqref{dj_pc_pb_dep} thus become
\begin{equation}\label{app_dj_mean}
	\Pr(\ket{\mathbf 0}) \approx \varepsilon(p_1)\left[1-\varepsilon(\bar p)\right]^{n-1},
\end{equation}
where $\varepsilon(p)$ is given by Eq.~\eqref{app_eps} and $\bar p$ is the mean error probability of the other $n-1$ qubits. In the Bernstein--Vazirani algorithm all qubits are equivalent and only their mean error probability matters, whereas here the error probability of the first qubit enters directly. Since $\varepsilon_1$ is small, the effect on the success probability of Eq.~\eqref{dj_avg} is nevertheless small. For example, for $n=30$, if the other qubits have $p_i=0.01$ and the first qubit has $p_1=0.015$, the success probability decreases from $0.866$ to $0.862$ (bit flip) and from $0.814$ to $0.810$ (depolarizing).

\end{appendices}

\bibliography{refs}


\begin{thebibliography}{44}
\ifx \bisbn   \undefined \def \bisbn  #1{ISBN #1}\fi
\ifx \binits  \undefined \def \binits#1{#1}\fi
\ifx \bauthor  \undefined \def \bauthor#1{#1}\fi
\ifx \batitle  \undefined \def \batitle#1{#1}\fi
\ifx \bjtitle  \undefined \def \bjtitle#1{#1}\fi
\ifx \bvolume  \undefined \def \bvolume#1{\textbf{#1}}\fi
\ifx \byear  \undefined \def \byear#1{#1}\fi
\ifx \bissue  \undefined \def \bissue#1{#1}\fi
\ifx \bfpage  \undefined \def \bfpage#1{#1}\fi
\ifx \blpage  \undefined \def \blpage #1{#1}\fi
\ifx \burl  \undefined \def \burl#1{\textsf{#1}}\fi
\ifx \doiurl  \undefined \def \doiurl#1{\url{https://doi.org/#1}}\fi
\ifx \betal  \undefined \def \betal{\textit{et al.}}\fi
\ifx \binstitute  \undefined \def \binstitute#1{#1}\fi
\ifx \binstitutionaled  \undefined \def \binstitutionaled#1{#1}\fi
\ifx \bctitle  \undefined \def \bctitle#1{#1}\fi
\ifx \beditor  \undefined \def \beditor#1{#1}\fi
\ifx \bpublisher  \undefined \def \bpublisher#1{#1}\fi
\ifx \bbtitle  \undefined \def \bbtitle#1{#1}\fi
\ifx \bedition  \undefined \def \bedition#1{#1}\fi
\ifx \bseriesno  \undefined \def \bseriesno#1{#1}\fi
\ifx \blocation  \undefined \def \blocation#1{#1}\fi
\ifx \bsertitle  \undefined \def \bsertitle#1{#1}\fi
\ifx \bsnm \undefined \def \bsnm#1{#1}\fi
\ifx \bsuffix \undefined \def \bsuffix#1{#1}\fi
\ifx \bparticle \undefined \def \bparticle#1{#1}\fi
\ifx \barticle \undefined \def \barticle#1{#1}\fi
\bibcommenthead
\ifx \bconfdate \undefined \def \bconfdate #1{#1}\fi
\ifx \botherref \undefined \def \botherref #1{#1}\fi
\ifx \url \undefined \def \url#1{\textsf{#1}}\fi
\ifx \bchapter \undefined \def \bchapter#1{#1}\fi
\ifx \bbook \undefined \def \bbook#1{#1}\fi
\ifx \bcomment \undefined \def \bcomment#1{#1}\fi
\ifx \oauthor \undefined \def \oauthor#1{#1}\fi
\ifx \citeauthoryear \undefined \def \citeauthoryear#1{#1}\fi
\ifx \endbibitem  \undefined \def \endbibitem {}\fi
\ifx \bconflocation  \undefined \def \bconflocation#1{#1}\fi
\ifx \arxivurl  \undefined \def \arxivurl#1{\textsf{#1}}\fi
\csname PreBibitemsHook\endcsname

\bibitem[\protect\citeauthoryear{Bernstein and Vazirani}{1997}]{bvalgo}
\begin{barticle}
\bauthor{\bsnm{Bernstein}, \binits{E.}},
\bauthor{\bsnm{Vazirani}, \binits{U.}}:
\batitle{Quantum complexity theory}.
\bjtitle{SIAM Journal on Computing}
\bvolume{26}(\bissue{5}),
\bfpage{1411}--\blpage{1473}
(\byear{1997})
\doiurl{10.1137/S0097539796300921}
\end{barticle}
\endbibitem

\bibitem[\protect\citeauthoryear{Deutsch and Jozsa}{1992}]{deutsch1992rapid}
\begin{barticle}
\bauthor{\bsnm{Deutsch}, \binits{D.}},
\bauthor{\bsnm{Jozsa}, \binits{R.}}:
\batitle{Rapid solution of problems by quantum computation}.
\bjtitle{Proceedings of the Royal Society of London. Series A: Mathematical and
  Physical Sciences}
\bvolume{439}(\bissue{1907}),
\bfpage{553}--\blpage{558}
(\byear{1992})
\doiurl{10.1098/rspa.1992.0167}
\end{barticle}
\endbibitem

\bibitem[\protect\citeauthoryear{Grover}{1997}]{grover}
\begin{barticle}
\bauthor{\bsnm{Grover}, \binits{L.K.}}:
\batitle{Quantum mechanics helps in searching for a needle in a haystack}.
\bjtitle{Phys. Rev. Lett.}
\bvolume{79},
\bfpage{325}--\blpage{328}
(\byear{1997})
\doiurl{10.1103/PhysRevLett.79.325}
\end{barticle}
\endbibitem

\bibitem[\protect\citeauthoryear{Shor}{1997}]{shor}
\begin{barticle}
\bauthor{\bsnm{Shor}, \binits{P.W.}}:
\batitle{Polynomial-time algorithms for prime factorization and discrete
  logarithms on a quantum computer}.
\bjtitle{SIAM Journal on Computing}
\bvolume{26}(\bissue{5}),
\bfpage{1484}--\blpage{1509}
(\byear{1997})
\doiurl{10.1137/S0097539795293172}
\end{barticle}
\endbibitem

\bibitem[\protect\citeauthoryear{Arute et~al.}{2019}]{arute}
\begin{barticle}
\bauthor{\bsnm{Arute}, \binits{F.}}, \betal:
\batitle{Quantum supremacy using a programmable superconducting processor}.
\bjtitle{Nature}
\bvolume{574}(\bissue{7779}),
\bfpage{505}--\blpage{510}
(\byear{2019})
\doiurl{10.1038/s41586-019-1666-5}
\end{barticle}
\endbibitem

\bibitem[\protect\citeauthoryear{Zhong et~al.}{2020}]{zhong}
\begin{barticle}
\bauthor{\bsnm{Zhong}, \binits{H.-S.}}, \betal:
\batitle{Quantum computational advantage using photons}.
\bjtitle{Science}
\bvolume{370}(\bissue{6523}),
\bfpage{1460}--\blpage{1463}
(\byear{2020})
\doiurl{10.1126/science.abe8770}
\end{barticle}
\endbibitem

\bibitem[\protect\citeauthoryear{Pokharel and Lidar}{2023}]{outperform}
\begin{barticle}
\bauthor{\bsnm{Pokharel}, \binits{B.}},
\bauthor{\bsnm{Lidar}, \binits{D.A.}}:
\batitle{Demonstration of algorithmic quantum speedup}.
\bjtitle{Phys. Rev. Lett.}
\bvolume{130},
\bfpage{210602}
(\byear{2023})
\doiurl{10.1103/PhysRevLett.130.210602}
\end{barticle}
\endbibitem

\bibitem[\protect\citeauthoryear{Kim et~al.}{2023}]{kim_utility}
\begin{barticle}
\bauthor{\bsnm{Kim}, \binits{Y.}},
\bauthor{\bsnm{Eddins}, \binits{A.}},
\bauthor{\bsnm{Anand}, \binits{S.}},
\bauthor{\bsnm{Wei}, \binits{K.X.}},
\bauthor{\bsnm{{van den Berg}}, \binits{E.}}, \betal:
\batitle{Evidence for the utility of quantum computing before fault tolerance}.
\bjtitle{Nature}
\bvolume{618}(\bissue{7965}),
\bfpage{500}--\blpage{505}
(\byear{2023})
\doiurl{10.1038/s41586-023-06096-3}
\end{barticle}
\endbibitem

\bibitem[\protect\citeauthoryear{Bravyi et~al.}{2018}]{bravyi2018}
\begin{barticle}
\bauthor{\bsnm{Bravyi}, \binits{S.}},
\bauthor{\bsnm{Gosset}, \binits{D.}},
\bauthor{\bsnm{K{\"o}nig}, \binits{R.}}:
\batitle{Quantum advantage with shallow circuits}.
\bjtitle{Science}
\bvolume{362}(\bissue{6412}),
\bfpage{308}--\blpage{311}
(\byear{2018})
\doiurl{10.1126/science.aar3106}
\end{barticle}
\endbibitem

\bibitem[\protect\citeauthoryear{Bravyi
  et~al.}{2020}]{bravyi2020NoisyShallowCircuits}
\begin{barticle}
\bauthor{\bsnm{Bravyi}, \binits{S.}},
\bauthor{\bsnm{Gosset}, \binits{D.}},
\bauthor{\bsnm{K{\"o}nig}, \binits{R.}},
\bauthor{\bsnm{Tomamichel}, \binits{M.}}:
\batitle{Quantum advantage with noisy shallow circuits}.
\bjtitle{Nat. Phys.}
\bvolume{16}(\bissue{10}),
\bfpage{1040}--\blpage{1045}
(\byear{2020})
\doiurl{10.1038/s41567-020-0948-z}
\end{barticle}
\endbibitem

\bibitem[\protect\citeauthoryear{Lou et~al.}{2026}]{lou_qucongan}
\begin{barticle}
\bauthor{\bsnm{Lou}, \binits{X.}},
\bauthor{\bsnm{Long}, \binits{C.}},
\bauthor{\bsnm{Wang}, \binits{L.}},
\bauthor{\bsnm{Wang}, \binits{Z.}}:
\batitle{{QuCon-GAN}: A parameter-efficient quantum-classical conditional {GAN}
  for handwritten digit generation}.
\bjtitle{IEEE Trans. Consum. Electron.}
(\byear{2026})
\doiurl{10.1109/TCE.2026.3708674}
\end{barticle}
\endbibitem

\bibitem[\protect\citeauthoryear{Liu and Lou}{2026}]{liu_hybrid}
\begin{barticle}
\bauthor{\bsnm{Liu}, \binits{H.}},
\bauthor{\bsnm{Lou}, \binits{X.}}:
\batitle{A parallel hybrid quantum-classical convolutional design using
  parameterized quantum circuits for image classification}.
\bjtitle{Quantum Eng.}
\bvolume{2026}(\bissue{1}),
\bfpage{6643049}
(\byear{2026})
\doiurl{10.1155/que2/6643049}
\end{barticle}
\endbibitem

\bibitem[\protect\citeauthoryear{Lou and Wang}{2025}]{lou_quanflex}
\begin{barticle}
\bauthor{\bsnm{Lou}, \binits{X.}},
\bauthor{\bsnm{Wang}, \binits{L.}}:
\batitle{{QuanFlex-GAN}: a flexible algorithm for image generation via quantum
  generative modeling}.
\bjtitle{Quantum Mach. Intell.}
\bvolume{7}(\bissue{2}),
\bfpage{87}
(\byear{2025})
\doiurl{10.1007/s42484-025-00316-x}
\end{barticle}
\endbibitem

\bibitem[\protect\citeauthoryear{Endo et~al.}{2021}]{endo2021Hybrid}
\begin{barticle}
\bauthor{\bsnm{Endo}, \binits{S.}},
\bauthor{\bsnm{Cai}, \binits{Z.}},
\bauthor{\bsnm{Benjamin}, \binits{S.C.}},
\bauthor{\bsnm{Yuan}, \binits{X.}}:
\batitle{Hybrid quantum-classical algorithms and quantum error mitigation}.
\bjtitle{Journal of the Physical Society of Japan}
\bvolume{90}(\bissue{3}),
\bfpage{032001}
(\byear{2021})
\doiurl{10.7566/JPSJ.90.032001}
\end{barticle}
\endbibitem

\bibitem[\protect\citeauthoryear{Preskill}{2018}]{preskill}
\begin{barticle}
\bauthor{\bsnm{Preskill}, \binits{J.}}:
\batitle{Quantum computing in the {NISQ} era and beyond}.
\bjtitle{Quantum}
\bvolume{2},
\bfpage{79}
(\byear{2018})
\doiurl{10.22331/q-2018-08-06-79}
\end{barticle}
\endbibitem

\bibitem[\protect\citeauthoryear{Krantz et~al.}{2019}]{krantz}
\begin{barticle}
\bauthor{\bsnm{Krantz}, \binits{P.}},
\bauthor{\bsnm{Kjaergaard}, \binits{M.}},
\bauthor{\bsnm{Yan}, \binits{F.}},
\bauthor{\bsnm{Orlando}, \binits{T.P.}},
\bauthor{\bsnm{Gustavsson}, \binits{S.}},
\bauthor{\bsnm{Oliver}, \binits{W.D.}}:
\batitle{A quantum engineer's guide to superconducting qubits}.
\bjtitle{Appl. Phys. Rev.}
\bvolume{6}(\bissue{2}),
\bfpage{021318}
(\byear{2019})
\doiurl{10.1063/1.5089550}
\end{barticle}
\endbibitem

\bibitem[\protect\citeauthoryear{Bruzewicz et~al.}{2019}]{bruzewicz}
\begin{barticle}
\bauthor{\bsnm{Bruzewicz}, \binits{C.D.}},
\bauthor{\bsnm{Chiaverini}, \binits{J.}},
\bauthor{\bsnm{McConnell}, \binits{R.}},
\bauthor{\bsnm{Sage}, \binits{J.M.}}:
\batitle{Trapped-ion quantum computing: Progress and challenges}.
\bjtitle{Appl. Phys. Rev.}
\bvolume{6}(\bissue{2}),
\bfpage{021314}
(\byear{2019})
\doiurl{10.1063/1.5088164}
\end{barticle}
\endbibitem

\bibitem[\protect\citeauthoryear{Gawron et~al.}{2012}]{groverinnoise}
\begin{barticle}
\bauthor{\bsnm{Gawron}, \binits{P.}},
\bauthor{\bsnm{Klamka}, \binits{J.}},
\bauthor{\bsnm{Winiarczyk}, \binits{R.}}:
\batitle{Noise effects in the quantum search algorithm from the viewpoint of
  computational complexity}.
\bjtitle{Int. J. Appl. Math. Comput. Sci.}
\bvolume{22}(\bissue{2}),
\bfpage{493}--\blpage{499}
(\byear{2012})
\doiurl{10.2478/v10006-012-0037-2}
\end{barticle}
\endbibitem

\bibitem[\protect\citeauthoryear{Faizan and
  Faryad}{2024}]{qpenoisefaizanfaryad}
\begin{bchapter}
\bauthor{\bsnm{Faizan}, \binits{M.}},
\bauthor{\bsnm{Faryad}, \binits{M.}}:
\bctitle{Simulation and analysis of quantum phase estimation algorithm in the
  presence of incoherent quantum noise channels}.
In: \beditor{\bsnm{Hemmer}, \binits{P.R.}},
\beditor{\bsnm{Migdall}, \binits{A.L.}} (eds.)
\bbtitle{Quantum Computing, Communication, and Simulation IV}.
\bsertitle{Proc. SPIE},
vol. \bseriesno{12911},
p. \bfpage{1291116}.
\bpublisher{SPIE},
\blocation{Bellingham, WA}
(\byear{2024}).
\doiurl{10.1117/12.2691149}
\end{bchapter}
\endbibitem

\bibitem[\protect\citeauthoryear{Gupta et~al.}{2023}]{effectofnoise}
\begin{botherref}
\oauthor{\bsnm{Gupta}, \binits{A.}},
\oauthor{\bsnm{Ghosh}, \binits{P.}},
\oauthor{\bsnm{Sen}, \binits{K.}},
\oauthor{\bsnm{Sen}, \binits{U.}}:
Effects of noise on performance of {B}ernstein--{V}azirani algorithm.
arXiv:2305.19745 [quant-ph]
(2023).
\doiurl{10.48550/arXiv.2305.19745}
\end{botherref}
\endbibitem

\bibitem[\protect\citeauthoryear{Cross et~al.}{2015}]{cross2015}
\begin{barticle}
\bauthor{\bsnm{Cross}, \binits{A.W.}},
\bauthor{\bsnm{Smith}, \binits{G.}},
\bauthor{\bsnm{Smolin}, \binits{J.A.}}:
\batitle{Quantum learning robust against noise}.
\bjtitle{Phys. Rev. A}
\bvolume{92}(\bissue{1}),
\bfpage{012327}
(\byear{2015})
\doiurl{10.1103/PhysRevA.92.012327}
\end{barticle}
\endbibitem

\bibitem[\protect\citeauthoryear{Rist{\`e} et~al.}{2017}]{riste2017}
\begin{barticle}
\bauthor{\bsnm{Rist{\`e}}, \binits{D.}},
\bauthor{\bsnm{{da Silva}}, \binits{M.P.}},
\bauthor{\bsnm{Ryan}, \binits{C.A.}}, \betal:
\batitle{Demonstration of quantum advantage in machine learning}.
\bjtitle{npj Quantum Inf.}
\bvolume{3},
\bfpage{16}
(\byear{2017})
\doiurl{10.1038/s41534-017-0017-3}
\end{barticle}
\endbibitem

\bibitem[\protect\citeauthoryear{Fern{\'a}ndez and
  Martin-Delgado}{2024}]{BVFernandez}
\begin{barticle}
\bauthor{\bsnm{Fern{\'a}ndez}, \binits{P.}},
\bauthor{\bsnm{Martin-Delgado}, \binits{M.A.}}:
\batitle{{Homomorphic encryption of the $k=2$ Bernstein--Vazirani algorithm}}.
\bjtitle{J. Phys. A}
\bvolume{57}(\bissue{36}),
\bfpage{365301}
(\byear{2024})
\doiurl{10.1088/1751-8121/ad6c04}
\end{barticle}
\endbibitem

\bibitem[\protect\citeauthoryear{Lubinski
  et~al.}{2023}]{lubinski2023Benchmarks}
\begin{barticle}
\bauthor{\bsnm{Lubinski}, \binits{T.}},
\bauthor{\bsnm{Johri}, \binits{S.}},
\bauthor{\bsnm{Varosy}, \binits{P.}},
\bauthor{\bsnm{Coleman}, \binits{J.}},
\bauthor{\bsnm{Zhao}, \binits{L.}},
\bauthor{\bsnm{Necaise}, \binits{J.}},
\bauthor{\bsnm{Baldwin}, \binits{C.H.}},
\bauthor{\bsnm{Mayer}, \binits{K.}},
\bauthor{\bsnm{Proctor}, \binits{T.}}:
\batitle{Application-oriented performance benchmarks for quantum computing}.
\bjtitle{IEEE Transactions on Quantum Engineering}
\bvolume{4},
\bfpage{1}--\blpage{32}
(\byear{2023})
\doiurl{10.1109/TQE.2023.3253761}
\end{barticle}
\endbibitem

\bibitem[\protect\citeauthoryear{Debnath
  et~al.}{2016}]{debnath2016Programmable}
\begin{barticle}
\bauthor{\bsnm{Debnath}, \binits{S.}},
\bauthor{\bsnm{Linke}, \binits{N.M.}},
\bauthor{\bsnm{Figgatt}, \binits{C.}},
\bauthor{\bsnm{Landsman}, \binits{K.A.}},
\bauthor{\bsnm{Wright}, \binits{K.}},
\bauthor{\bsnm{Monroe}, \binits{C.}}:
\batitle{Demonstration of a small programmable quantum computer with atomic
  qubits}.
\bjtitle{Nature}
\bvolume{536}(\bissue{7614}),
\bfpage{63}--\blpage{66}
(\byear{2016})
\doiurl{10.1038/nature18648}
\end{barticle}
\endbibitem

\bibitem[\protect\citeauthoryear{Linke et~al.}{2017}]{linke2017Comparison}
\begin{barticle}
\bauthor{\bsnm{Linke}, \binits{N.M.}},
\bauthor{\bsnm{Maslov}, \binits{D.}},
\bauthor{\bsnm{Roetteler}, \binits{M.}},
\bauthor{\bsnm{Debnath}, \binits{S.}},
\bauthor{\bsnm{Figgatt}, \binits{C.}},
\bauthor{\bsnm{Landsman}, \binits{K.A.}},
\bauthor{\bsnm{Wright}, \binits{K.}},
\bauthor{\bsnm{Monroe}, \binits{C.}}:
\batitle{Experimental comparison of two quantum computing architectures}.
\bjtitle{Proceedings of the National Academy of Sciences}
\bvolume{114}(\bissue{13}),
\bfpage{3305}--\blpage{3310}
(\byear{2017})
\doiurl{10.1073/pnas.1618020114}
\end{barticle}
\endbibitem

\bibitem[\protect\citeauthoryear{Blinov
  et~al.}{2021}]{blinov2021CloudComparison}
\begin{barticle}
\bauthor{\bsnm{Blinov}, \binits{S.}},
\bauthor{\bsnm{Wu}, \binits{B.}},
\bauthor{\bsnm{Monroe}, \binits{C.}}:
\batitle{Comparison of cloud-based ion trap and superconducting quantum
  computer architectures}.
\bjtitle{AVS Quantum Science}
\bvolume{3}(\bissue{3}),
\bfpage{033801}
(\byear{2021})
\doiurl{10.1116/5.0058187}
\end{barticle}
\endbibitem

\bibitem[\protect\citeauthoryear{Wright et~al.}{2019}]{wright}
\begin{barticle}
\bauthor{\bsnm{Wright}, \binits{K.}},
\bauthor{\bsnm{Beck}, \binits{K.M.}},
\bauthor{\bsnm{Debnath}, \binits{S.}}, \betal:
\batitle{Benchmarking an 11-qubit quantum computer}.
\bjtitle{Nat. Commun.}
\bvolume{10},
\bfpage{5464}
(\byear{2019})
\doiurl{10.1038/s41467-019-13534-2}
\end{barticle}
\endbibitem

\bibitem[\protect\citeauthoryear{Johansson and Larsson}{2019}]{qsimmdpi}
\begin{barticle}
\bauthor{\bsnm{Johansson}, \binits{N.}},
\bauthor{\bsnm{Larsson}, \binits{J.-{\AA}.}}:
\batitle{Quantum simulation logic, oracles, and the quantum advantage}.
\bjtitle{Entropy}
\bvolume{21}(\bissue{8}),
\bfpage{800}
(\byear{2019})
\doiurl{10.3390/e21080800}
\end{barticle}
\endbibitem

\bibitem[\protect\citeauthoryear{Johansson and Larsson}{2017}]{djcontrast}
\begin{barticle}
\bauthor{\bsnm{Johansson}, \binits{N.}},
\bauthor{\bsnm{Larsson}, \binits{J.-{\AA}.}}:
\batitle{Efficient classical simulation of the {D}eutsch--{J}ozsa and {S}imon's
  algorithms}.
\bjtitle{Quantum Information Processing}
\bvolume{16}(\bissue{9}),
\bfpage{233}
(\byear{2017})
\doiurl{10.1007/s11128-017-1679-7}
\end{barticle}
\endbibitem

\bibitem[\protect\citeauthoryear{Orrell and Loer}{2021}]{orrell2021}
\begin{barticle}
\bauthor{\bsnm{Orrell}, \binits{J.L.}},
\bauthor{\bsnm{Loer}, \binits{B.}}:
\batitle{Sensor-assisted fault mitigation in quantum computation}.
\bjtitle{Phys. Rev. Appl.}
\bvolume{16}(\bissue{2}),
\bfpage{024025}
(\byear{2021})
\doiurl{10.1103/PhysRevApplied.16.024025}
\end{barticle}
\endbibitem

\bibitem[\protect\citeauthoryear{Urbanek et~al.}{2021}]{mitigdep}
\begin{barticle}
\bauthor{\bsnm{Urbanek}, \binits{M.}},
\bauthor{\bsnm{Nachman}, \binits{B.}},
\bauthor{\bsnm{Pascuzzi}, \binits{V.R.}},
\bauthor{\bsnm{He}, \binits{A.}},
\bauthor{\bsnm{Bauer}, \binits{C.W.}},
\bauthor{\bsnm{{de Jong}}, \binits{W.A.}}:
\batitle{Mitigating depolarizing noise on quantum computers with
  noise-estimation circuits}.
\bjtitle{Phys. Rev. Lett.}
\bvolume{127},
\bfpage{270502}
(\byear{2021})
\doiurl{10.1103/PhysRevLett.127.270502}
\end{barticle}
\endbibitem

\bibitem[\protect\citeauthoryear{Magesan et~al.}{2011}]{magesan}
\begin{barticle}
\bauthor{\bsnm{Magesan}, \binits{E.}},
\bauthor{\bsnm{Gambetta}, \binits{J.M.}},
\bauthor{\bsnm{Emerson}, \binits{J.}}:
\batitle{Scalable and robust randomized benchmarking of quantum processes}.
\bjtitle{Phys. Rev. Lett.}
\bvolume{106},
\bfpage{180504}
(\byear{2011})
\doiurl{10.1103/PhysRevLett.106.180504}
\end{barticle}
\endbibitem

\bibitem[\protect\citeauthoryear{Vovrosh et~al.}{2021}]{vovrosh}
\begin{barticle}
\bauthor{\bsnm{Vovrosh}, \binits{J.}},
\bauthor{\bsnm{Khosla}, \binits{K.E.}},
\bauthor{\bsnm{Greenaway}, \binits{S.}},
\bauthor{\bsnm{Self}, \binits{C.}},
\bauthor{\bsnm{Kim}, \binits{M.S.}},
\bauthor{\bsnm{Knolle}, \binits{J.}}:
\batitle{Simple mitigation of global depolarizing errors in quantum
  simulations}.
\bjtitle{Phys. Rev. E}
\bvolume{104},
\bfpage{035309}
(\byear{2021})
\doiurl{10.1103/PhysRevE.104.035309}
\end{barticle}
\endbibitem

\bibitem[\protect\citeauthoryear{{Google Quantum AI}}{2021}]{google2021}
\begin{barticle}
\bauthor{\bsnm{{Google Quantum AI}}}:
\batitle{Exponential suppression of bit or phase errors with cyclic error
  correction}.
\bjtitle{Nature}
\bvolume{595}(\bissue{7867}),
\bfpage{383}--\blpage{387}
(\byear{2021})
\doiurl{10.1038/s41586-021-03588-y}
\end{barticle}
\endbibitem

\bibitem[\protect\citeauthoryear{Nielsen and Chuang}{2010}]{neilson}
\begin{bbook}
\bauthor{\bsnm{Nielsen}, \binits{M.A.}},
\bauthor{\bsnm{Chuang}, \binits{I.L.}}:
\bbtitle{Quantum Computation and Quantum Information},
\bedition{10th anniversary} edn.
\bpublisher{Cambridge University Press},
\blocation{Cambridge}
(\byear{2010}).
\doiurl{10.1017/CBO9780511976667}
\end{bbook}
\endbibitem

\bibitem[\protect\citeauthoryear{Basit et~al.}{2017}]{Basit_2017}
\begin{barticle}
\bauthor{\bsnm{Basit}, \binits{A.}},
\bauthor{\bsnm{Badshah}, \binits{F.}},
\bauthor{\bsnm{Ali}, \binits{H.}},
\bauthor{\bsnm{Ge}, \binits{G.-Q.}}:
\batitle{Protecting quantum coherence and discord from decoherence of
  depolarizing noise via weak measurement and measurement reversal}.
\bjtitle{Europhysics Letters}
\bvolume{118}(\bissue{3}),
\bfpage{30002}
(\byear{2017})
\doiurl{10.1209/0295-5075/118/30002}
\end{barticle}
\endbibitem

\bibitem[\protect\citeauthoryear{McEwen et~al.}{2022}]{mcewen}
\begin{barticle}
\bauthor{\bsnm{McEwen}, \binits{M.}},
\bauthor{\bsnm{Faoro}, \binits{L.}},
\bauthor{\bsnm{Arya}, \binits{K.}}, \betal:
\batitle{Resolving catastrophic error bursts from cosmic rays in large arrays
  of superconducting qubits}.
\bjtitle{Nat. Phys.}
\bvolume{18}(\bissue{1}),
\bfpage{107}--\blpage{111}
(\byear{2022})
\doiurl{10.1038/s41567-021-01432-8}
\end{barticle}
\endbibitem

\bibitem[\protect\citeauthoryear{Wilen et~al.}{2021}]{wilen}
\begin{barticle}
\bauthor{\bsnm{Wilen}, \binits{C.D.}},
\bauthor{\bsnm{Abdullah}, \binits{S.}},
\bauthor{\bsnm{Kurinsky}, \binits{N.A.}}, \betal:
\batitle{Correlated charge noise and relaxation errors in superconducting
  qubits}.
\bjtitle{Nature}
\bvolume{594}(\bissue{7863}),
\bfpage{369}--\blpage{373}
(\byear{2021})
\doiurl{10.1038/s41586-021-03557-5}
\end{barticle}
\endbibitem

\bibitem[\protect\citeauthoryear{Harper et~al.}{2020}]{harper}
\begin{barticle}
\bauthor{\bsnm{Harper}, \binits{R.}},
\bauthor{\bsnm{Flammia}, \binits{S.T.}},
\bauthor{\bsnm{Wallman}, \binits{J.J.}}:
\batitle{Efficient learning of quantum noise}.
\bjtitle{Nat. Phys.}
\bvolume{16}(\bissue{12}),
\bfpage{1184}--\blpage{1188}
(\byear{2020})
\doiurl{10.1038/s41567-020-0992-8}
\end{barticle}
\endbibitem

\bibitem[\protect\citeauthoryear{{de Vega} and Alonso}{2017}]{devega}
\begin{barticle}
\bauthor{\bsnm{{de Vega}}, \binits{I.}},
\bauthor{\bsnm{Alonso}, \binits{D.}}:
\batitle{Dynamics of non-{Markovian} open quantum systems}.
\bjtitle{Rev. Mod. Phys.}
\bvolume{89}(\bissue{1}),
\bfpage{015001}
(\byear{2017})
\doiurl{10.1103/RevModPhys.89.015001}
\end{barticle}
\endbibitem

\bibitem[\protect\citeauthoryear{Wallman and Emerson}{2016}]{wallman}
\begin{barticle}
\bauthor{\bsnm{Wallman}, \binits{J.J.}},
\bauthor{\bsnm{Emerson}, \binits{J.}}:
\batitle{Noise tailoring for scalable quantum computation via randomized
  compiling}.
\bjtitle{Phys. Rev. A}
\bvolume{94},
\bfpage{052325}
(\byear{2016})
\doiurl{10.1103/PhysRevA.94.052325}
\end{barticle}
\endbibitem

\bibitem[\protect\citeauthoryear{Ballance et~al.}{2016}]{ballance2016}
\begin{barticle}
\bauthor{\bsnm{Ballance}, \binits{C.J.}},
\bauthor{\bsnm{Harty}, \binits{T.P.}},
\bauthor{\bsnm{Linke}, \binits{N.M.}},
\bauthor{\bsnm{Sepiol}, \binits{M.A.}},
\bauthor{\bsnm{Lucas}, \binits{D.M.}}:
\batitle{High-fidelity quantum logic gates using trapped-ion hyperfine qubits}.
\bjtitle{Phys. Rev. Lett.}
\bvolume{117}(\bissue{6}),
\bfpage{060504}
(\byear{2016})
\doiurl{10.1103/PhysRevLett.117.060504}
\end{barticle}
\endbibitem

\bibitem[\protect\citeauthoryear{Foxen et~al.}{2020}]{foxen2020}
\begin{barticle}
\bauthor{\bsnm{Foxen}, \binits{B.}},
\bauthor{\bsnm{Neill}, \binits{C.}},
\bauthor{\bsnm{Dunsworth}, \binits{A.}}, \betal:
\batitle{Demonstrating a continuous set of two-qubit gates for near-term
  quantum algorithms}.
\bjtitle{Phys. Rev. Lett.}
\bvolume{125}(\bissue{12}),
\bfpage{120504}
(\byear{2020})
\doiurl{10.1103/PhysRevLett.125.120504}
\end{barticle}
\endbibitem

\end{thebibliography}

\end{document}